\documentclass[lettersize,journal]{IEEEtran}

\usepackage{cite}
\usepackage{amsmath,amssymb,amsfonts}
\usepackage{graphicx}
\usepackage{listings}
\usepackage{xcolor}
\usepackage{url}
\usepackage{xurl}
\usepackage{hyperref}
\usepackage{booktabs}
\usepackage{array}
\usepackage{enumitem}
\usepackage{pifont}
\usepackage{tcolorbox}
\usepackage{subcaption}
\usepackage{multirow}
\usepackage{xspace}
\usepackage{makecell}
\usepackage{comment}
\usepackage{orcidlink}
\def\BibTeX{{\rm B\kern-.05em{\sc i\kern-.025em b}\kern-.08em
    T\kern-.1667em\lower.7ex\hbox{E}\kern-.125emX}}

\newcommand{\mypar}[1]{\smallskip\noindent\textbf{#1.}\xspace}

\begin{document}

\title{The Infinite Mutation Engine? Measuring Polymorphism in LLM-Generated Offensive Code}

\author{Gabriel Hortea \orcidlink{0009-0003-8971-915X},
        Mario Varlese \orcidlink{0009-0006-1981-3548},
        Simon Pietro Romano \orcidlink{0000-0002-5876-0382},
        Juan Tapiador \orcidlink{0000-0002-4573-3967}
\thanks{Gabriel Hortea and Juan Tapiador are with the Department of Computer Science, Universidad Carlos III de Madrid, 28911 Leganés, Spain (e-mail: bhortea@pa.uc3m.es; jestevez@inf.uc3m.es).}
\thanks{Mario Varlese and Simon Pietro Romano are with the Department of Electrical Engineering and Information Technology, University of Naples Federico II, 80125 Naples, Italy (e-mail: mario.varlese@unina.it; spromano@unina.it).}
}

\maketitle

\begin{abstract}
Malware authors traditionally rely on polymorphic techniques to produce variants within a malware family, complicating detection. Generative AI enables attackers to automatically synthesize structurally diverse payloads with identical behavior, raising the question of how much evasion Large Language Models (LLMs) actually provide. Recent work has assumed that LLMs can produce highly polymorphic payloads, leaving unquantified the variation that emerges when an attacker repeatedly generates a payload, or explicitly instructs the model to avoid prior implementations. In this work, we address this gap by measuring the polymorphic capacity of a commercial model (Claude Opus~4.6) acting as an automated malware generator. We build a dual-agent, four-stage pipeline that generates, tests, and refines a data-exfiltration payload comprising file traversal, encryption, exfiltration, and integration. We produce payloads under two settings: (a) functional prompting, and (b) explicit prompting injecting a structured history of prior variants. Evaluating structural (TEDs) and semantic (embeddings) variance, we find that explicit prompting fragments the structural space, creating variants that are hard to cluster for some ML classifiers. This structural diversity is operationally cheap, costing \$0.41--\$0.73 per validated payload. However, static n-gram analysis reveals that because the LLM treats inter-module boundaries as immutable, a single static signature targeting them achieves 100\% detection across all generations. Finally, a complementary evaluation of seven local small language models (SLMs) (20M--1.5B parameters) confirms that current open-weights architectures lack the generative stability required for reliable polymorphic synthesis.
\end{abstract}


% Import sections
\section{Introduction}
\label{sec:intro}

\IEEEPARstart{L}{LMs} are increasingly being used in software engineering to automate code generation, assist in runtime debugging, and optimize complex software pipelines~\cite{ashrafi2025enhancing, amazon2025selfdebug}. This capability also lowers the technical barrier for malicious actors by enabling rapid synthesis of functional logic without requiring deep programming expertise~\cite{unit42024maliciousllms}. Recent work shows that attackers can manipulate commercial LLMs to bypass safety guardrails and generate evasive malware variants capable of degrading the detection rates of both commercial antivirus engines and machine-learning classifiers~\cite{li2025llmmalware, llmalmorph2025}. In practice, this threat is no longer theoretical: threat actors are actively integrating LLM-assisted generation into their attack chains to scale deployment, build advanced evasion techniques, and automate exploitation campaigns~\cite{anthropic2025espionage, khadgi2025lamehug, layerx2025aimalware}.

A key potential advantage that LLMs provide to attackers is their capacity for rapid, automated code mutation. Traditional polymorphic and metamorphic malware relies on rigid, human-authored packing routines or obfuscation templates to alter its representation and structure with the aim of evading signature-based detection~\cite{tripwire2023polymorphic, brezinski2023metamorphicsurvey}. LLMs, by contrast, naturally introduce syntactic and structural diversity each time they generate code. When tasked with producing malicious payloads, LLM-based frameworks can output functionally identical but structurally distinct implementations~\cite{cognicrypt2026}. Industry studies confirm that this AI-driven polymorphism\footnote{Malware research has traditionally distinguished between polymorphic code, which changes only its appearance by encrypting the same payload under mutating keys, and metamorphic code, which rewrites its code and changes its structure. In this paper, we use the term \emph{polymorphism} to refer to the different code implementations produced by an LLM from a specification, even when these outputs could potentially have different structural properties and be more akin to metamorphic code.} enables attackers to evade both static signature rules (e.g., YARA) and similarity-based heuristic clustering employed by Endpoint Detection and Response (EDR) systems~\cite{blueflux2025, muhr2026slopoly, gen2026promptmorphism}.

Despite this growing adoption, the underlying mechanics of LLM-driven polymorphism remain poorly understood. Existing work has demonstrated the evasion capabilities of iteratively obfuscated LLM malware~\cite{unit42024llmsjavascript}, explored multistep prompt-based malware delivery~\cite{brodt2026promptwarekillchainprompt}, and proposed LLM-assisted frameworks for malware classification~\cite{qian2025lamd}. However, these studies treat the generative model as a black box, evaluating the final evasiveness of the code without measuring the generative variation itself. Specifically, the field lacks (i) a rigorous quantification of how much variation naturally occurs when an LLM repeatedly performs the exact same malicious objective, and (ii) to what extent deliberate prompting strategies amplify this divergence. To address this gap, we formulate two primary research questions:
\begin{description}
  \item[RQ1:] When an LLM is repeatedly invoked with an identical functional task specification, with no mention of polymorphism, how much code variation naturally emerges across the population of generated implementations?
  \item[RQ2:] If the LLM is explicitly instructed to vary its code structure while preserving behavior, and a history of prior implementations is injected to force novelty, does the degree and distribution of polymorphism increase?
\end{description}

\mypar{Contributions}
In this work, we present the first systematic measurement study quantifying the polymorphism of LLM-synthesized offensive code. We focus on a realistic end-to-end attack simulation divided into four sequential stages: file traversal, payload encryption, data exfiltration, and integration. We then evaluate the generated code under two prompting strategies (inherent natural diversity vs.\ explicit history-driven novelty) and along two orthogonal axes: structural distance via Zhang-Shasha Tree Edit Distance (TED) \cite{zhangshasha}, and semantic distance via code-specific UniXcoder embeddings \cite{guo2022unixcoder}. Specifically, we make the following contributions:
\begin{itemize}[leftmargin=*]
    \item We design and implement a dual-agent orchestration framework that autonomously generates, tests, and refines multi-stage malware-like payloads (\S\ref{sec:methodology}). By pairing generator LLMs with tester LLMs alongside deterministic Python validation, we guarantee that all measured polymorphism is extracted exclusively from functionally correct code, removing the noise of LLM hallucinations.
    
    \item We demonstrate that a commercial LLM (Claude Opus~4.6) acts as a polymorphic engine even when provided only with strictly functional prompts (\S\ref{sec:results-rq1}), observing baseline divergence: mean structural distances reach $0.24$--$0.38$ for active stages, while semantic distances remain low ($\leq$0.24). This confirms the model produces diverse surface-level implementations without altering high-level behavior.

    \item We introduce a history-injection prompting mechanism that explicitly forces the LLM to structurally diverge from prior generations (\S\ref{sec:results-rq2}), generating $\geq$88\% noise in density-based clustering across three stages. This structural obfuscation collapses the semantic space ($K{=}1$ in three of four stages), demonstrating that the LLM successfully hides distinct algorithmic variations from ML-based representations (\S\ref{sec:clustering}).
    
    \item We present a static n-gram analysis proving that current LLM polymorphism is \emph{wide but shallow} (\S\ref{sec:clustering-ngram}), where despite successfully defeating semantic clustering, the LLM treats inter-module API contracts as immutable scaffolding. A single static signature rule targeting these integration seams achieves 100\% detection across the entire cohort.

    \item We quantify the search effort required for this LLM-driven polymorphism, finding it economical (\S\ref{sec:results-costs}). While explicit mode increases average token consumption by roughly $5\times$, it demands only a negligible increase in generation attempts. An attacker can produce fully validated payloads at an effective API cost of just \$0.41 (inherent) to \$0.73 (explicit).

    \item We evaluate seven local open-weight SLMs (20M--1.5B parameters)
    on the file traversal stage, showing that only the largest model produces a non-trivial yield ($5.1\%$ over 2,480 attempts) and its structural diversity reflects generative instability rather than controlled mutation (\S\ref{sec:results-slms}).

    \item Finally, we analyze the defensive implications and asymmetries of LLM-driven polymorphism (\S\ref{sec:discussion}), showing that while invariant inter-module contracts preserve traditional static detection, the minimal cost and semantic collapse create risks for future monolithic payloads.
\end{itemize}

\mypar{Research Artifact}
To support open science and reproducibility, all prompt templates, pipeline configurations, and measurement scripts are publicly available at:
\begin{center}
    \url{https://anonymous.4open.science/r/llm_polymorphism-C0D0}
\end{center}

\section{Related Work}
\label{sec:related}
This section discusses prior work on classical malware polymorphism and metamorphism, and the use of LLMs for offensive operations and code generation.

\mypar{Polymorphic and Metamorphic Malware}
Polymorphic malware changes its representation while preserving behavior, making signature-based detection difficult~\cite{tripwire2023polymorphic}. Metamorphic malware goes further by rewriting its own code on each generation to evade static analysis~\cite{fortinet2016metamorphic}. Survey work by Masabo \emph{et} al.\ and Brezinski \emph{et} al.\ ~\cite{masabo2017polymorphicsurvey,brezinski2023metamorphicsurvey} concludes that static analysis is efficient but fragile under obfuscation, whereas dynamic analysis is more robust but costly to deploy at scale. More recent work explores multi-stage and ensemble detection architectures to better handle polymorphic families~\cite{chaikovskyi2024comprehensive}.

\mypar{LLM-Driven Offensive Operations in the Wild}
Anthropic describes what appears to be the first reported
large-scale cyber espionage campaign in which an AI agent performed the
bulk of reconnaissance, exploitation, and data exfiltration with
limited human intervention~\cite{anthropic2025espionage}. Google's GTIG
documents multiple threat actors using AI tools across the attack
lifecycle, including families such as \emph{PROMPTFLUX} and
\emph{PROMPTSTEAL}~\cite{google2025gtig}. Microsoft warns that LLMs
lower barriers to attacks and argues for AI-augmented
defense to counter AI-enabled offense~\cite{microsoft2024ageofai}. GuardSix reports
that APT28 deployed \emph{LAMEHUG}, the first AI-generated
malware used by a nation-state actor in an active
campaign~\cite{khadgi2025lamehug}. Bitdefender documents APT36
adopting vibeware, AI-malware generated at
scale to lower development cost and increase
evasion~\cite{tudorica2026vibeware}. ESET's analysis of
\emph{PromptSpy} extends this threat to the Android ecosystem~\cite{eset2026promptspy}. Complementary reports
such as PromptIntel~\cite{promptintel2025nova} and LOLMIL~\cite{dreadnode2025lolmil} document how adversarial prompts and
``living off the land'' models are being cataloged and
shared. The broader
threat landscape for AI systems is formalized in MITRE ATLAS~\cite{mitreATLAS2024}.

\mypar{LLM-Generated Polymorphic Code}
Industry reports such as IBM X-Force's \emph{Slopoly} and Gen Digital's \emph{Promptmorphism} show that LLMs can be prompted to produce structurally distinct ransomware and loader variants that evade static signatures and similarity-based clustering~\cite{muhr2026slopoly,gen2026promptmorphism}, while Crytica Security argues that future AI-generated malware will be preemptively polymorphic~\cite{crytica2023polymorphic}. At the code level, Unit~42 show that LLM-driven obfuscation of malicious JavaScript can increase evasion rates~\cite{unit42024llmsjavascript}. Li \emph{et al.} embed LLMs in a reinforcement-learning loop to generate adversarial Windows malware that bypasses MalConv~\cite{li2025llmmalware}. Raz \emph{et al.} and Brodt \emph{et al.} describe self-composing ransomware and prompt-based kill chains that regenerate their own code~\cite{ransomware3point0_2025,brodt2026promptwarekillchainprompt}. Lin \emph{et al.} introduce \emph{Code as a Weapon}, using LLMs and MITRE ATT\&CK–style prompts to generate end-to-end malware, assessing its detectability against commercial antivirus engines~\cite{lin2025codeweapon}. Madani proposes a self-testing mutation framework where LLMs rewrite program subroutines while unit tests enforce semantic equivalence~\cite{madani2024metamorphic}. Carvalho \emph{et al.} studied the detectability of LLM-generated malware~\cite{carvalho2025generatingmalware}, Saha and Shukla's MalGEN the multi-agent framework~\cite{saha2025malgen}, and Akil \emph{et al.}'s LLMalMorph variant generator~\cite{llmalmorph2025} demonstrate that LLM-based pipelines can synthesize realistic, evasive samples. Whereas prior work focuses on feasibility and evasion rates, we treat polymorphism as the primary object of study, quantitatively characterizing its structural and semantic diversity.

\section{Threat Model}
\label{sec:threat_model}
In this work, we focus on an evolving threat paradigm where adversaries use LLMs to automate the generation of structurally diverse malware payloads. We frame our threat model around a Man-At-The-End (MATE) scenario~\cite{collberg1997taxonomy,falcarin11,llvm2025mate}, explicitly adapting it to the context of generative AI. While classical MATE literature typically describes scenarios where an attacker targets a fixed piece of software they already possess~\cite{basile2019mate}, we extend this paradigm to automated payload synthesis. In our generative MATE adaptation, rather than possessing the target software, the attacker possesses an unconstrained code-generation environment (the ``AI code factory''~\cite{cyberinst2025factory}). The attacker has full control to iteratively test, mutate, and validate the code without defender visibility, while the defender later controls the execution endpoint and attempts to detect the resulting payloads using static signatures.

\mypar{Attacker Profile, Capabilities and Objectives}
We model an adversary seeking to deploy offensive payloads at scale, such as localized ransomware or data exfiltration implants. While traditional polymorphic and metamorphic malware require human-authored mutation engines to alter the code representation~\cite{tripwire2023polymorphic}, we study an attacker who delegates code synthesis entirely to an LLM~\cite{pedarla2025rise} and expects the produced code to be polymorphic. We assume that the attacker has access to commercial LLM APIs and is capable of executing prompt-engineering techniques such as role-play framing, context laundering, and iterative refinement~\cite{secnora2026manipulation, unit42024maliciousllms}. Operationally, the attacker utilizes an offline orchestrator to generate and test the code. By establishing a local feedback loop (such as AEG~\cite{brumley2011aeg}), the attacker ensures that the LLM's outputs are functionally correct prior to deployment, avoiding broken payloads~\cite{cognicrypt2026}. The attacker's primary objective is to maximize differences in the generated malware code while maintaining the core semantic behavior of the payload, aiming to defeat static signature generation (e.g., YARA rule matching) and heuristic clustering (such as n-gram analysis) typically employed by modern EDR systems~\cite{pedarla2025rise, cognicrypt2026, payer2014maldiv}. 

\mypar{Defender Capabilities} 
We model a defender equipped with state-of-the-art static analysis tools and code-similarity metrics designed to group polymorphic variants into known malware families ~\cite{bayer2009scalable,runwal2012opcode,gao2013discriminant}. The defender has full visibility into the malicious payloads. We assume that the defender does not initially possess cryptographic hashes or any other rigid signature for the newly generated LLM payloads, as each generated instance acts as a structurally unique sample~\cite{cyberinst2025factory}.

\mypar{Scope}
We limit the scope of our threat model to static code analysis and explicitly exclude evasion of dynamic (behavioral) analysis from our evaluation. While the generated malware payloads will inevitably produce a runtime behavioral footprint (e.g., initiating connections or executing file I/O operations), bypassing dynamic analysis is orthogonal to our focus~\cite{bayer2009scalable,gysel2024eagleeye,adaptive2025polymorphic}. A successful attack in our framework is defined strictly by the pipeline's ability to automatically generate, validate, and output a functionally correct payload that is ``sufficiently'' different from previous samples.

\section{Methodology}
\label{sec:methodology}
\begin{figure*}[t!]
	\centering
	\includegraphics[width=\textwidth]{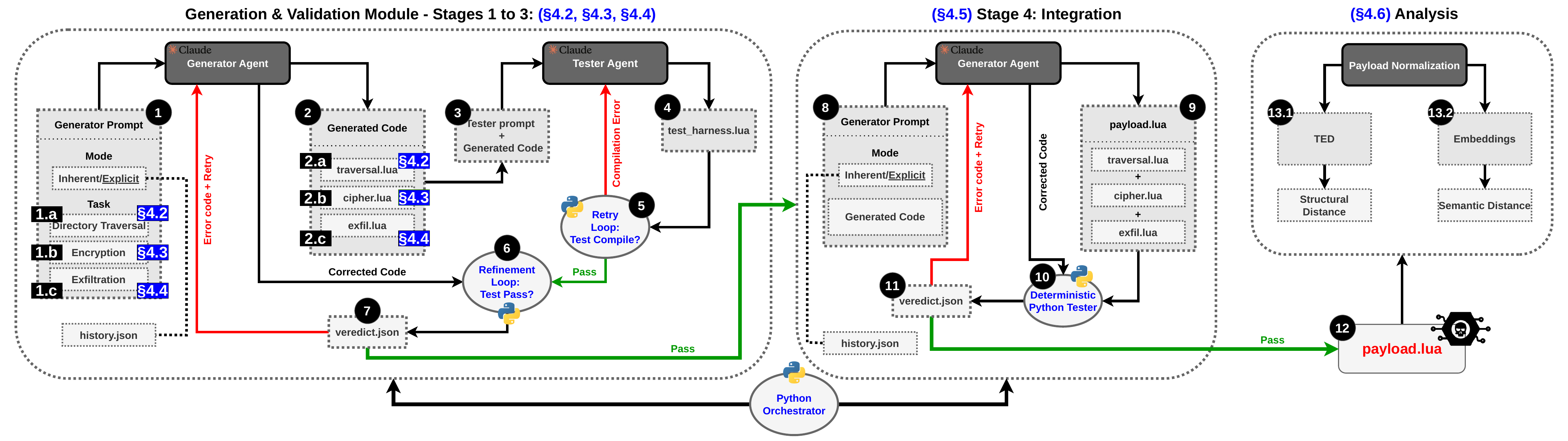}
	\caption{Overview of the methodology pipeline.}
	\label{fig:pipeline}
\end{figure*}

This section details the experimental design, prompt architecture, and implementation details of our methodology.
Figure~\ref{fig:pipeline} provides an overview of the pipeline, organized into three phases: a shared generation \& validation module for stages~1--3 (left, steps~\textcircled{\small 1}--\textcircled{\small 7}), the integration stage (center, steps~\textcircled{\small 8}--\textcircled{\small 11}), and the analysis stage (right, steps~\textcircled{\small 12}--\textcircled{\small 13}). Following this structure, \S\ref{subsec:methodology_overview} outlines the core objectives, language constraints, and the multi-agent paradigm that drives all three phases; \S\ref{subsec:stage1_traversal}--\S\ref{subsec:stage3_exfil} details the traversal, encryption, and exfiltration stages; \S\ref{subsec:stage4_integration} covers the integration stage; and \S\ref{subsec:stage5_analysis} describes the analysis phase.

\subsection{Overview}
\label{subsec:methodology_overview}
To measure the polymorphic capabilities of LLMs, we built an experimental framework capable of autonomously generating, verifying, and refining the required task implementations.

\mypar{The Attack Simulation Task}
The task is to synthesize a four-stage data exfiltration payload, mimicking the standard operational lifecycle of modern ransomware and data-theft implants~\cite{ransomware3point0_2025}: 
(1) Traversal: locating target files,
(2) Encryption: securing the files via custom cryptographic routines,
(3) Exfiltration: transmitting the ciphertexts over a command and control channel, and
(4) Integration: stitching the modules into an orchestrated executable.
By segmenting the payload, we force the model to handle programmatic challenges (e.g., filesystem I/O and network sockets), ensuring our measurements capture a broad spectrum of structural and algorithmic adaptations.
We do not model the initial access vector, delivery mechanisms, or privilege escalation techniques.

\mypar{Language Selection}
We selected Lua~\cite{lua} for three reasons. First, as an interpreted language, the raw script itself is the deployed artifact; avoiding compiler optimization passes and exposing source-level polymorphism to static signature tools. Second, its minimalist standard library forces the LLM to write custom, low-level logic rather than relying on standard wrappers, maximizing the search space for structural variation. Finally, Lua is increasingly prevalent in modern threats, where adversaries bundle lightweight, embedded runtimes to execute stealthy, multi-platform payloads on endpoints lacking native interpreters~\cite{mitre2024lua}.

\mypar{The Generator-Tester Architecture} 
LLMs are susceptible to hallucinations and errors when generating complex code~\cite{arora2024promptdesign}. To avoid measuring non-viable implementations, we employ a dual-agent paradigm powered by Claude Opus 4.6 \cite{claude_opus}. Within the Generation \& Validation module (Figure~\ref{fig:pipeline}, left), step~\textcircled{\small 1} configures the stage-specific generator prompt, step~\textcircled{\small 2} emits the corresponding module, and steps~\textcircled{\small 3}--\textcircled{\small 7} implement the tester-driven refinement loop shared by Stages~1--3. 
\begin{itemize}[leftmargin=*]
    \item \textbf{Generator agents:} For each of the four stages, we instantiate an LLM with a clean context window to generate the target Lua module. To bypass safety filters, agents are prompted with: (a) persona adoption ~\cite{unit42024multi, onsecurity2025jailbreak}, framing the model as an academic assistant in an authorized experiment; and (b) cognitive bounding, commanding it to output exclusively valid Lua code within Markdown backticks while suppressing conversational filler, apologies, and disclaimers.

    \item \textbf{Tester agents:} Testers receive raw generator output and write custom Lua test harnesses to validate behavioral integrity. This LLM testing applies strictly to Stages~1--3 (steps~\textcircled{\small 3}--\textcircled{\small 6} in Figure~\ref{fig:pipeline}). Stage~4 instead uses a deterministic Python script (step~\textcircled{\small 10}); because individual modules are already verified, validating the payload requires only standard execution and output parsing, saving token overhead.
\end{itemize}

A deterministic Python orchestrator (bottom of Figure~\ref{fig:pipeline}) manages the state machine between the agents, feeding errors back to the generator and ensuring that only structurally viable and logically correct code advances from phase to phase.

\mypar{Experimental Study Modes} 
To isolate the causes of code variation, generator agents are prompted using two distinct study modes across all four stages:
\begin{itemize}[leftmargin=*]
    \item \textbf{Inherent mode (baseline):} The LLM receives only functional constraints without mention of structural diversity. This establishes a baseline for the natural drift caused by the autoregressive decoding process of the model (RQ1).
    \item \textbf{Explicit mode (prompted):} The LLM is commanded to mutate its implementation (RQ2). After each successful run, the orchestrator caches extracted polymorphic comment blocks in \texttt{history.json} file (Fig \ref{fig:history_json}). In subsequent iterations, this history is injected into generator prompts for Stages~1--3 (step~\textcircled{\small 1}) and for integration (step~\textcircled{\small 8}). More details of this study mode can be found in Appendix \ref{app:explicit_mode}.
\end{itemize}
Although higher temperature or nucleus sampling in Claude Opus 4.6 increases variation, we maintain default API parameters to isolate prompting effects and support reproducibility. Altering these settings degrades alignment, inflating hallucinations and breaking code correctness~\cite{arora2024promptdesign}.

\subsection{Stage 1: Traversal}
\label{subsec:stage1_traversal}
The pipeline initiates with traversal, simulating the malware's initial reconnaissance phase~\cite{reconnaissance}. In Figure~\ref{fig:pipeline}, the traversal sub-task is encoded as part of the shared generator prompt in step~\textcircled{\small 1.a}, and the resulting \texttt{traversal.lua} module appears in step~\textcircled{\small 2.a}.
The goal is to synthesize a Lua module that navigates a local filesystem, identifies target files by extension, and evades designated decoys. More details can be found in Appendix \ref{app:prompt_traversal}.

\mypar{Generator Agent Task and Constraints} 
The generator is prompted to write a self-contained \texttt{traversal.lua} module restricted to standard Lua 5.4 libraries (\texttt{io}, \texttt{os}, \texttt{string}). It must recursively traverse subdirectories, return an empty table if no files match, and handle permission errors. In inherent mode, receiving only functional constraints establishes a natural drift baseline. In explicit mode, the prompt is injected with historical evolutionary context (\texttt{history.json}, Fig. \ref{fig:history_json}) and instructed to mutate code along at least two of three algorithmic axes: traversal strategy, pattern matching, or discovery mode.

\mypar{Tester Agent and the Validation Sandbox} 
Upon receiving the code (step~\textcircled{\small 2.a}), the traversal tester (step~\textcircled{\small 3}) writes a \texttt{test\_harness.lua} script (step~\textcircled{\small 4}). The orchestrator provisions an isolated sandbox containing nested target files alongside identically named decoys in restricted directories. Given the absolute paths for targets (\texttt{EXPECTED\_FILES}) and decoys (\texttt{DECOY\_FILES}), the tester generates a harness executing six baseline assertions (Table \ref{tab:traversal_tests}).

\mypar{Orchestration and Refinement Loops}
Two nested refinement loops govern agent-sandbox interaction (steps~\textcircled{\small 5}--\textcircled{\small 6}, Fig.~\ref{fig:pipeline}). Executing the test harness produces a serialized \texttt{verdict.json} state object (step~\textcircled{\small 7}, Fig.\ref{fig:verdict_stage1}). If the harness contains syntax errors, runtime stack traces are returned to the tester in a harness retry loop. If the module fails an assertion, the specific error is extracted from \texttt{verdict.json} and fed back to the generator in a logic refinement loop. Exceeding maximum retry limits triggers a session wipe and full regeneration. Once all assertions are met, the discovered file paths are stored in Python memory as deterministic input for Stage~2.

\subsection{Stage 2: Encryption}
\label{subsec:stage2_cipher}
Following traversal, the pipeline transitions to the encryption stage, simulating the cryptographic phase of ransomware or secure exfiltration payloads (step~\textcircled{\small 1.b}) and its generated module \texttt{cipher.lua} ~\textcircled{\small 2.b} in Fig.~\ref{fig:pipeline}). The objective is to synthesize a mathematically invertible symmetric cipher without relying on external libraries. More details can be found in Section \ref{app:prompt_encryption} of the Appendix.

\mypar{Generator Agent Task and Constraints} 
The generator is prompted to produce a self-contained module called \texttt{cipher.lua}. Using the academic persona established in Section~\ref{subsec:methodology_overview}, the agent must output strictly binary-safe code (handling \texttt{0x00} null bytes) relying exclusively on Lua's native bitwise (\texttt{\textasciitilde}, \texttt{\&}, \texttt{|}, \texttt{<<}, \texttt{>>}) and arithmetic operators. In explicit mode, the generator mutates code along three structural axes: algorithm family, key handling, and processing mode. Additionally, Stage~1 file metadata (e.g., sizes, extensions) is used to induce data-dependent polymorphism.

\mypar{Tester Agent and Mathematical Validation} 
Stage 2 validation requires strict mathematical proof of invertibility. The orchestrator injects a fixed test key and the Stage~1 file paths into the tester's prompt. The tester synthesizes a Lua harness executing five baseline assertions to verify deterministic, lossless encryption (Table \ref{tab:cipher_tests}).

\mypar{Orchestration and Context Propagation} 
The execution follows the refinement logic of steps~\textcircled{\small 5}--\textcircled{\small 7} (Fig.~\ref{fig:pipeline}), logging results to a \texttt{verdict.json} state object (Fig.\ref{fig:verdict_stage1}). Syntax errors (e.g., failing to open files in binary mode via \texttt{"rb"}) trigger the harness retry loop, while logical failures (e.g., string truncation in \texttt{[T1]}) trigger the logic refinement loop to rewrite the mathematical routine. Once all assertions are met, the \texttt{cipher.lua} is cached in memory for the exfiltration stage.

\subsection{Stage 3: Exfiltration}
\label{subsec:stage3_exfil}
Following encryption, the pipeline advances to exfiltration, synthesizing a module to securely transmit encrypted buffers to a Command and Control (C2) endpoint (step~\textcircled{\small 1.c} and its \texttt{exfil.lua} module~\textcircled{\small 2.c} in Fig.~\ref{fig:pipeline}). More details can be found in Section~\ref{app:prompt_exfil} of the Appendix.

\mypar{Generator Agent Task and Constraints} 
The generator produces an \texttt{exfil.lua} module. As Lua lacks native networking primitives, this is the sole stage where importing an external dependency (\texttt{luasocket}) is authorized. Constraints require binary-safe transmission and graceful failure handling (returning \texttt{false} without thread crashes). In explicit mode, the evolution context commands the agent to mutate across four network axes: transport protocol, encoding scheme, send strategy, and connection state.

\mypar{Tester Agent and Mock Infrastructure} 
To test unverified socket code without internet exposure, the orchestrator provisions a localized mock C2 listener (bound to \texttt{127.0.0.1} on a dynamic port) matching the chosen transport protocol. Injecting this host, port, and a test payload into the tester prompt forces the agent to synthesize a Lua harness executing three local network assertions (Table~\ref{tab:exfil_tests}). While the tester verifies Lua-side execution (steps~\textcircled{\small 3}--\textcircled{\small 6}), the orchestrator performs an offline check (\texttt{[T4]}) on the mock C2 buffer to ensure uncorrupted socket delivery.

\mypar{Orchestration and Context Propagation} 
Harness execution follows standard refinement loops (steps~\textcircled{\small 5}--\textcircled{\small 7}, Fig.~\ref{fig:pipeline}), logging to a \texttt{verdict.json} object (Fig.~\ref{fig:verdict_stage1}). If the module passes Lua-side checks (\texttt{[T1]}--\texttt{[T3]}) but fails the mock C2 byte check (\texttt{[T4]}), the orchestrator intercepts the failure and feeds the error back to the generator to rebuild the encoding logic. Passing all four assertions caches \texttt{exfil.lua} and its network parameters, advancing the state machine to integration.

\subsection{Stage 4: Integration}
\label{subsec:stage4_integration}
The final integration phase synthesizes a unified \texttt{payload.lua} script to sequentially import and orchestrate the validated Stage 1--3 outputs into a single execution flow (center block in Fig.~\ref{fig:pipeline}, steps~\textcircled{\small 8}--\textcircled{\small 11}). More details can be found in Section~\ref{app:prompt_integration} of the Appendix.

\mypar{Generator Agent Task and Constraints} 
The generator (step~\textcircled{\small 8}) writes a master script loading the validated \texttt{traversal.lua}, \texttt{cipher.lua}, and \texttt{exfil.lua} modules via \texttt{dofile()}. The script invokes traversal to discover targets, iterate through each file in binary mode, encrypt buffers, and transmit them via exfiltration. It enforces a mandatory 0.2-second \texttt{socket.sleep()} between transmissions to ensure mock C2 stability during multi-file bursts. To enable deterministic validation, the script prints standardized stdout logs (e.g., \texttt{SENT: <filepath>} and \texttt{DONE: <N> sent, <M> failed}). In explicit mode, the evolution context directs the generator to mutate the control flow along three architectural axes: data buffering, processing flow, and error handling (producing \texttt{payload.lua}, step~\textcircled{\small 9}, Fig.~\ref{fig:pipeline}).

\mypar{Python Validation (asymmetric testing)} 
Because individual module behavior is already proven, re-invoking an LLM to evaluate the stitched payload adds unnecessary token latency and API overhead. Instead, the orchestrator performs offline, deterministic validation (step~\textcircled{\small 10}). Provisioning a final sandbox containing the test folder, validated modules, and a multi-connection mock C2 instance, the orchestrator executes \texttt{payload.lua} directly, regex-parses stdout strings, and serializes the state into a Stage~4 \texttt{verdict.json} (step~\textcircled{\small 11}).

\mypar{End-to-End Orchestration} 
The final logic refinement loop runs strictly between the orchestrator and the integration generator. The orchestrator compares \texttt{SENT} logs against Stage~1's \texttt{found\_files} array, retrieves raw bytes caught by the C2, decrypts them with \texttt{cipher.lua}, and performs byte-for-byte verification against original sandbox files. This state is serialized into an expanded \texttt{verdict.json} object (Fig.~\ref{fig:verdict_stage4}) mapping received payloads to filesystem origins. If scope errors (e.g., missing global references across stitched modules) or decryption roundtrips fail, Python stack traces are returned to the generator for refinement. Passing this end-to-end test concludes code generation, saving the validated \texttt{payload.lua} (step~\textcircled{\small 12}) to disk for analysis.

\subsection{Analysis}
\label{subsec:stage5_analysis}
The final phase of the methodology is the analysis stage (right block in Figure~\ref{fig:pipeline}), where the orchestrator extracts validated \texttt{payload.lua} artifacts (step~\textcircled{\small 12}) and normalizes them into canonical representations. We extract both a structural view via Tree Edit Distance (TED; step~\textcircled{\small 13.1}) and a semantic view via code embeddings (step~\textcircled{\small 13.2}) to quantify polymorphism for Section~\ref{sec:results}.

\mypar{Payload Normalization}
To ensure metrics reflect programmatic logic, modules undergo a normalization pass: regex strips injected \texttt{[[POLYMORPHISM]]} tracking blocks and single-line Lua comments, followed by trimming line whitespace and discarding empty lines.

\mypar{TED Extraction and Structural Distance}
For structural analysis (step~\textcircled{\small 13.1}), we compute pairwise Zhang-Shasha Tree Edit Distance (TED)~\cite{zhangshasha} between payload ASTs using the APTED algorithm~\cite{pawlik2015rted} via the \texttt{edist} library~\cite{Paassen2015EDM}. Raw edit distance is normalized by the larger tree's node count, yielding a metric bounded in $[0.0, 1.0]$.

\mypar{Embedding and Semantic Distance}
For semantic analysis (step~\textcircled{\small 13.2}), normalized payloads are embedded using \texttt{microsoft/unixcoder-base}~\cite{guo2022unixcoder}. Scripts exceeding context limits are chunked into 1,500-character blocks, embedded independently, and mean-pooled into a single vector representation. We then compute pairwise cosine distances across all payload vectors, bounded in $[0.0, 1.0]$.

\section{Results}
\label{sec:results}
We executed the complete pipeline across 200 independent runs, generated 100 under the inherent mode and 100 using the explicit mode. First, we establish the natural baseline of code variation by examining the inherent (\S\ref{sec:results-rq1}, RQ1) and the explicit (\S\ref{sec:results-rq2}, RQ2) mode. Then, we study the effect of both types of LLM polymorphism on malware classification via clustering (\S\ref{sec:clustering}) and n-grams (\S\ref{sec:clustering-ngram}). Next, we discuss the operational effort and the monetary cost required to execute the pipeline (\S\ref{sec:results-costs}). Finally, we analyze polymorphism in smaller models (\S\ref{sec:results-slms}).

\begin{figure*}[t!]
  \centering
  \includegraphics[width=\textwidth]{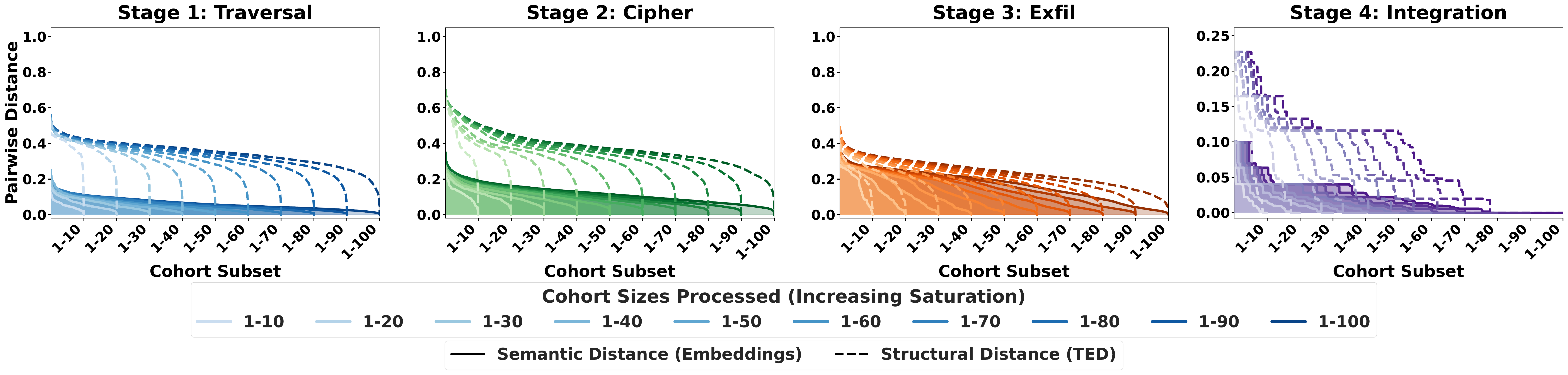}
  \caption{Cumulative diversity profiles using the structural (TED) and semantic (embeddings) distances. Darker curves correspond to larger cohorts (up to all 100 generations).}
  \label{fig:cumulative-both}
\end{figure*}

\subsection{Baseline Inherent Polymorphism}
\label{sec:results-rq1}
\begin{figure}[t!]
  \centering
  \includegraphics[width=\columnwidth]{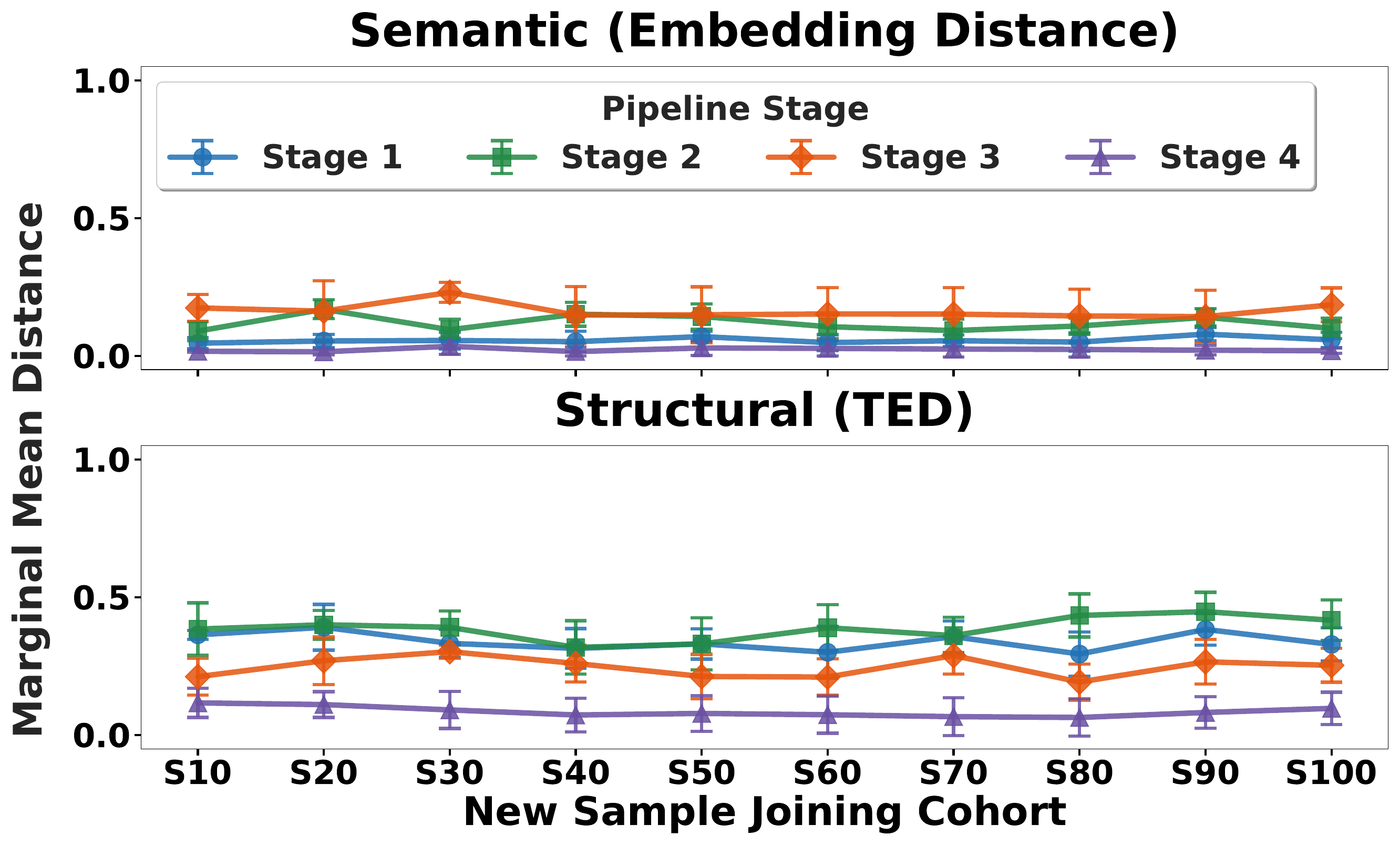}
  \caption{Evolution of marginal mean distances between each new sample and all previously generated samples.}
  \label{fig:diversity-evolution}
\end{figure}

To evaluate natural baseline polymorphism (RQ1) following \S\ref{subsec:stage5_analysis}, we construct $100\times 100$ pairwise distance matrices per stage along two orthogonal axes: (i)~structural distance (normalized TEDs~\cite{zhangshasha}) and (ii)~semantic distance (UniXcoder cosine distance~\cite{guo2022unixcoder}). We analyze these data through two lenses: cumulative diversity profiles (Fig.~\ref{fig:cumulative-both}), showing distance distributions across cohort subsets ($k\in\{10,20,\dots,100\}$) as population size grows; and diversity evolution (Fig.~\ref{fig:diversity-evolution}), tracking the marginal mean distance between each new sample and all prior samples $\{1,\dots,s-1\}$ to test for structural collapse over time. Qualitative source code inspection complements these quantitative metrics.

\begin{figure*}[t!]
  \centering
  \includegraphics[width=\textwidth]{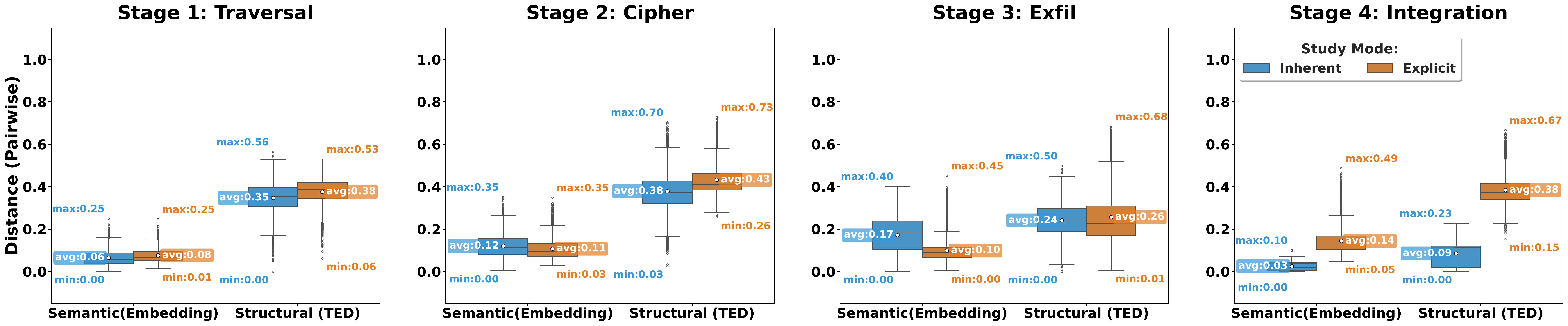}
 \caption{Comparative polymorphism distribution for the Inherent
  (blue) and Explicit (orange) modes. Each subplot shows pairwise
  semantic (left) and structural (right) distance distributions for
  one pipeline stage. Numerical annotations mark the empirical
  minimum, mean, and maximum over all off-diagonal pairs in the
  corresponding $100\times 100$ distance matrix.}
  \label{fig:polymorphism-boxplots}
\end{figure*}

\mypar{Stage 1: Traversal}
The structural curves (Figure~\ref{fig:cumulative-both}) center around a mean distance of~0.35, with the full cohort ($k=100$) spanning distances from near zero to a maximum of~0.56. The semantic distances remain low, with a mean of~0.065. Manual inspection reveals the source of this pattern: the LLM produces three distinct traversal strategies: (a) recursive DFS (60 samples), (b) iterative stack-based DFS (40 samples), and (c) within the iterative family, a queue-based BFS variant in 6 of those 40. Yet all implementations converge on the same underlying primitives: \texttt{io.popen}-based directory listing with a glob-to-Lua pattern converter, cross-platform branching between \texttt{ls~-1A} and \texttt{dir /b}, and \texttt{pcall} error handling. The algorithmic split and varying helper-function layouts (explicit basename extraction in 37\%, path normalization in 30\%, dedicated \texttt{entry\_type()} dispatchers in 7\%) drive the moderate structural distance, while the shared functional core explains the low semantic distance. Furthermore, Figure~\ref{fig:diversity-evolution} shows the marginal structural mean remains stable across the entire cohort ($0.40\to0.34$), with no sign of collapse. Notably, the model introduces safety features absent from the prompt in a minority of samples: depth limiting (23\%), symlink avoidance (9\%), and result sorting (17\%).

\mypar{Stage 2: Cipher}
The cipher stage exhibits the highest pairwise distances in the inherent mode, with a mean structural distance of~0.38 and a maximum reaching~0.70. The model also records the highest size coefficient of variation across stages (31.1\%), with implementations ranging from 60 to 305 lines. Manual classification reveals five distinct cipher architectures. The dominant family (89 out of 100) consists of Feistel networks, split between hybrid designs that reuse RC4's Key Scheduling Algorithm for key expansion (56 samples) and pure Feistel designs with custom hash-based key derivation (33 samples), while nine implement pure RC4 stream ciphers. Two outliers stand alone: an additive PRNG stream cipher that uses modular addition instead of XOR, making the encrypt and decrypt paths asymmetric; and a hybrid that derives Feistel round keys through an RC4 KSA and applies the resulting block cipher in CBC mode with a deterministically derived IV. Within the Feistel family, further variation arises from different block modes (stream-mode keystream generation in the majority, CBC in~18, CTR in~8), round counts (16 in 64 samples vs.\ 8 in~9), and padding strategies (length headers in 58 vs.\ PKCS7 in 12). Correspondingly, the semantic mean distance of~0.12 is the highest among all stages, reflecting the genuine algorithmic diversity. In Figure~\ref{fig:diversity-evolution}, the marginal structural distance oscillates within a stable band ($0.43\to0.36$) rather than decaying, confirming sustained diversity. Remarkably, 22 samples discard the first 768 bytes of RC4 keystream output to mitigate known statistical biases, a textbook hardening technique the model applies without being prompted~\cite{mironov2002not}.

\mypar{Stage 3: Exfiltration}
The exfiltration module sits in a middle ground of structural diversity, with a mean structural distance of~0.24 and a maximum of~0.50. The marginal structural distance in Figure~\ref{fig:diversity-evolution} exhibits a baseline of~0.24 across the full cohort. The semantic distance, with a mean of~0.17, is higher than either the traversal or cipher stages, indicating that the code-specific embedding model captures genuine functional differences among the exfiltration implementations despite their shared transport layer. All 100 samples converge on TCP transport via LuaSocket, hex encoding for binary safety, a SYNC-based wire protocol, and a fixed 10-second timeout. The structural variation that does exist stems entirely from wire format: 79 samples frame messages with an explicit \texttt{END} marker while 21 rely on length-only framing; 27 embed a sample identifier in the wire header; and the length field itself is encoded as 8-character hex (33 samples), 16-character hex (18), or decimal (8). Only 3 samples (3\%) introduce retry logic on connection failure, and 25\% implement multi-frame decoders for parsing concatenated messages.

\mypar{Stage 4: Integration}
In Figure~\ref{fig:cumulative-both}, all structural curves cluster near the low end of the distance scale, with a mean of just~0.085 and a maximum of~0.23. Semantic distances essentially collapse near zero with a mean of~0.025 and a maximum of only~0.10. Figure~\ref{fig:diversity-evolution} reinforces this, showing flat, near-zero marginal distances across all 100 samples. Code inspection confirms that all 100 samples use identical \texttt{dofile()}-based module loading, with exactly the same \texttt{socket.sleep(0.2)} inter-file delay, and identical status messaging (\texttt{SENT}/\texttt{FAILED}/\texttt{DONE}). The max/min byte ratio is only 1.4$\times$, and the size coefficient of variation (7.8\%) is four times lower than any other stage. The only two measurable variations are the presence of argument validation with \texttt{os.exit} (7 samples) and conditional sleep skipping after the last file (29 samples). Because the integration prompt tasks the LLM with stitching previously validated modules together, the model defaults to a single template, exhibiting negligible polymorphism at this stage.

\subsection{Explicit Polymorphism}
\label{sec:results-rq2}
We now discuss the impact of explicit polymorphism prompting and
history injection (RQ2). We reuse the
distance matrices described in \S\ref{sec:results-rq1} and
compare the full distributions between the
inherent and explicit cohorts. Figure~\ref{fig:polymorphism-boxplots} 
reports, for each stage and
metric, the distribution of the pairwise distances as boxplots
split by study mode. As with the inherent analysis, we complement
these metrics with qualitative inspection of the generated source
code and the structured polymorphism records maintained in the
history buffer.

\mypar{Stage 1: Traversal}
Explicit prompting amplifies both structural and semantic
polymorphism in the traversal module. The mean structural distance
increases to 0.38, and the minimum pairwise distance rises 
to~0.06, indicating that explicit prompting eliminates near-duplicate
implementations. On the semantic axis, the mean distance increases
to~0.076, confirming that while the explicit mode
generates more structurally varied code, the functional behaviour
remains similarly constrained. Qualitative inspection helps explain
this pattern: BFS via a queue table becomes the most common strategy
(37\%), followed by iterative DFS with an explicit stack (36\%)
and recursive DFS (27\%). The model also introduces an entirely new
structural strategy: 52\% of traversal implementations now use
coroutine-based directory listing, wrapping \texttt{io.popen} output
in Lua coroutines. Similarly, the three file-matching
strategies--glob-to-Lua-pattern conversion (33\%), suffix extraction
via \texttt{string.sub} (33\%), and exact string comparison with
manual extension parsing (34\%)-- are now evenly distributed.
Cross-platform branching drops to 31\%, as explicit prompting forces the
model away from repetition. Conversely, the emergent safety features
observed in the inherent mode disappear under explicit prompting.

\mypar{Stage 2: Cipher}
Cipher exhibits the highest structural polymorphism in both modes,
and explicit prompting pushes it further. The mean structural
distance increases to~0.43, and the maximum reaches~0.73.
The structural minimum jumps to~0.26, indicating that explicit
prompting eliminates near-duplicate cipher pairs entirely (every
implementation is at least 25\% structurally distinct from every
other). The size coefficient of variation increases to
38.0\%, with implementations ranging from 102 to 600 lines. 
On the semantic axis, the mean distance decreases marginally to~0.11,
reflecting the nature of the code-specific embedding model:
while each explicit cipher employs a unique algorithm (Table~\ref{tab:crypto_architectures}), they all conform to the same
\texttt{encrypt}/\texttt{decrypt} API contract, which the embedding
model correctly identifies as functionally equivalent. Under explicit
prompting, every implementation uses a globally unique
cryptographic architecture. Only 31\% of explicit implementations
still mention a Feistel structure, and just 1\% retains RC4; the
remaining 68\% employ architectures not present in the inherent
cohort. Block modes also diversify: CBC appears in 26 samples, CTR in~11, 
while sponge-like absorption/squeeze modes and OFB feedback variants emerge as new
categories. Notably, the explicit prompt's data-seeding mechanism
causes each cipher to derive its structural parameters from the
specific file metadata of the target host, producing
instance-specific configurations.

\begin{table}[t!]
  \centering
  \caption{Cryptographic architectures generated in Stage 2 under explicit prompting.}
  \label{tab:crypto_architectures}
  \small
  \begin{tabular*}{\linewidth}{@{\extracolsep{\fill}}lp{0.65\linewidth}@{}}
    \toprule
    \textbf{Family} & \textbf{Referenced Algorithms \& Structures} \\ 
    \midrule
    SPN & SKINNY, PRINCE, Whirlpool, Twofish \\
    \addlinespace
    Stream (LFSR) & A5/1, Grain-v2, WAGE \\
    \addlinespace
    ARX & ChaCha, Salsa20, Threefish, SipHash, SPARKLE \\
    \addlinespace
    Sponge & Ascon, Keccak, Gimli, Xoodoo, NORX \\
    \addlinespace
    Feistel / Legacy & Feistel variants, RC4 \\
    \bottomrule
  \end{tabular*}
\end{table}

\mypar{Stage 3: Exfiltration}
For exfiltration, explicit prompting widens the structural
range rather than shifting the centre. The mean structural distance
increases to~0.26, and the maximum climbs to~0.68. In contrast, 
the mean semantic distance decreases to~0.10. This reversal has a concrete
explanation: explicit prompting causes 92\% of samples to adopt
base64 encoding and 99\% to use chunked transmission, creating a functionally uniform
core that the code-specific embedding model identifies as
semantically similar. Manual inspection reveals 9 different combinations of
exfiltration strategies, defined by four axes: transport protocol,
data encoding, send granularity, and connection pattern. The dominant
combination (54\%) uses TCP with base64 encoding, 512-byte chunks,
and a per-chunk reconnection pattern. The second-largest cluster
(32\%) switches to UDP with base64 and 480-byte chunks, introducing
a transport bifurcation not observed in the inherent mode. 
Five samples retain the inherent-style pattern
(TCP/hex/single-connection), three samples use UDP with a persistent
connection, and three further outliers employ custom encodings: XOR
obfuscation layered atop base64, XOR with hex encoding, and one
sample using HTTP as transport. 

\mypar{Stage 4: Integration}
The largest effect of explicit prompting appears in the integration
stage. The mean semantic distance increases nearly
six-fold to~$\approx 0.14$ (max~0.49), and the mean structural
distance leaps to~$\approx 0.38$ with a maximum of~0.67. The
structural minimum also rises to~0.15, confirming that
every pair of explicit integration scripts is at least 15\%
structurally distinct. Code inspection reveals a near-complete
structural inversion. The size coefficient of variation rises
to 23.6\%, with files spanning 93--246 lines (vs.\ 53--67
in the inherent mode). While all 100 samples still use
\texttt{dofile()}-based loading, the model now generates diversity
along three orthogonal dimensions summarized in
Table~\ref{tab:integration_strategies}.

\subsection{Cluster-based Diversity Analysis}
\label{sec:clustering}
To evaluate practical evasion against unsupervised clustering, we apply DBSCAN~\cite{ester1996density} (\texttt{scikit-learn}~\cite{pedregosa2011scikit}, \texttt{metric=\textquotesingle precomputed\textquotesingle}, $\mathtt{min\_samples}{=}3$) across all 16 distance matrices. DBSCAN is selected because it accepts arbitrary precomputed distance matrices, determines cluster count $K$ dynamically, and explicitly isolates unclustered noise points. To avoid scale bias across varying stage/mode distributions, we execute a per-matrix parameter sweep, selecting $\varepsilon \in [0.05, 0.90]$ (step~$0.025$) that maximizes the mean Silhouette score~\cite{rousseeuw1987silhouettes} for $K{\ge}2$. If no $\varepsilon$ yields $K{\ge}2$, we fall back to the $k$-NN elbow heuristic ($k{=}3$)~\cite{ester1996density}, reflecting unimodal distribution. Table~\ref{tab:clustering} summarizes all clustering results; Section~\ref{app:umap} of the Appendix provides corresponding 2-D UMAP projections.

\begin{table}[t!]
  \centering
  \caption{Integration strategies generated in Stage 4 under explicit prompting.}
  \label{tab:integration_strategies}
  \small
  \begin{tabular*}{\linewidth}{@{\extracolsep{\fill}}lp{0.65\linewidth}@{}}
    \toprule
    \textbf{Dimension} & \textbf{Generated Implementations} \\ 
    \midrule
    \multirow{3}{*}{\parbox{2.5cm}{Buffering\\(99\% Unique)}} & Byte-level (XOR, nibble swap, Gray code) \\
    & Matrix permutation (spiral, Latin square) \\
    & Re-encryption (TEA, RC4 mask, Vigenère) \\
    \addlinespace
    \multirow{3}{*}{\parbox{2.5cm}{Process Order\\(97\% Unique)}} & Schröder, Lucas, Pell, Padovan \\
    & Jacobsthal, tribonacci, super Catalan \\
    & Divisor-count ranking \\
    \addlinespace
    \multirow{2}{*}{\parbox{2.5cm}{Error Handling}} & Fail-fast via \texttt{os.exit} (50\%) \\
    & Resilient \texttt{pcall} wrapper (34\%) \\
    \bottomrule
  \end{tabular*}
\end{table}

\begin{table}[t!]
\centering
\caption{Cluster-based diversity metrics for different modes and stages.
  Per-cell values are: $K$ is
  the number of non-noise clusters; $\mathrm{Sil}{\in}[-1,1]$ is the mean
  Silhouette score; $DB$ is the Davies-Bouldin
  index \cite{davies1979cluster}; $\bar{d}$ is the mean pairwise
  distance; and $\varepsilon$ is the DBSCAN neighborhood radius
  selected by the silhouette-maximizing sweep. Unimodal cohorts ($K{=}1$) are marked with the
  ${\dagger}$~symbol and have undefined Silhouette and $DB$ values, with
  $\bar{d}$ serving as the only quality indicator.}
\label{tab:clustering}
\renewcommand{\arraystretch}{1.25}
\setlength{\tabcolsep}{3pt}
\footnotesize
\resizebox{\columnwidth}{!}{%
\begin{tabular}{@{}ll cc cc@{}}
\toprule
\multirow{2}{*}{\textbf{Mode}}
  & \multirow{2}{*}{\textbf{Stage}}
  & \multicolumn{2}{c}{\textbf{TED}}
  & \multicolumn{2}{c}{\textbf{Embedding}} \\
\cmidrule(lr){3-4}\cmidrule(lr){5-6}
  &
  & \makecell{$(K,\mathrm{Sil},DB,\bar{d})$}
  & $\varepsilon$
  & \makecell{$(K,\mathrm{Sil},DB,\bar{d})$}
  & $\varepsilon$ \\
\midrule
\multirow{4}{*}{I}
  & S1
    & $(2,0.443,0.657,0.347)$  & $0.100$
    & $(2,0.503,0.684,0.065)$  & $0.050$ \\
  & S2
    & $(2,0.673,0.358,0.379)$  & $0.125$
    & $(3,0.449,0.525,0.120)$  & $0.050$ \\
  & S3
    & $(3,0.574,0.575,0.241)$  & $0.050$
    & $(3,0.475,1.435,0.171)$  & $0.050$ \\
  & S4
    & $\mathbf{(3,0.826,0.228,0.085)}$ & $0.050$
    & $\mathbf{(1^\dagger,{-},{-},0.025)}$ & $0.050$ \\
\midrule
\multirow{4}{*}{E}
  & S1
    & $(2,0.720,0.283,0.376)$  & $0.125$
    & $\mathbf{(1^\dagger,{-},{-},0.076)}$ & $0.097$ \\
  & S2
    & $(2,0.483,0.549,0.432)$  & $0.300$
    & $\mathbf{(1^\dagger,{-},{-},0.108)}$ & $0.120$ \\
  & S3
    & $(2,0.592,0.571,0.257)$  & $0.350$
    & $\mathbf{(1^\dagger,{-},{-},0.099)}$ & $0.110$ \\
  & S4
    & $(2,0.251,0.784,0.384)$  & $0.225$
    & $(2,0.384,0.683,0.144)$  & $0.075$ \\
\bottomrule
\end{tabular}%
}
\end{table}

\mypar{Stage~1: Traversal}
The traversal stage exhibits high structural dispersion in both
modes. DBSCAN identifies only $K_{\text{TED}}{=}2$ structural
clusters in each mode, but with noise ratios of 91\% (inherent) and
93\% (explicit). The vast majority of traversal payloads are thus
structural outliers with no dense neighborhood, pointing to
near-unitary diversity: each script is structurally unique,
consistent with the observation
in~\S\ref{sec:results-rq1}. In the embedding space, inherent mode
organises traversal into $K_{\text{Emb}}{=}2$ semantic families
($\mathrm{Sil}{=}0.503$, $DB{=}0.684$, $\bar{d}{=}0.065$,
$\varepsilon{=}0.050$), reflecting the primary algorithmic split
between recursive and iterative traversal strategies documented
in~\S\ref{sec:results-rq1}. Under explicit mode, the embedding space
collapses to a single unimodal cluster ($K{=}1$) with
$\bar{d}_{\text{Emb}}{=}0.076$: no $\varepsilon$ in the sweep
achieves $K{\geq}2$, indicating that the code-specific embedding
model views all 100 explicit traversal variants as functionally
equivalent despite their structural diversity.

\mypar{Stage~2: Cipher}
Explicit mode pushes structural diversity: at
$\bar{d}{=}0.432$, DBSCAN requires $\varepsilon{=}0.300$ to form
even two coarse clusters, which retain 88\% noise and exhibit
moderate separation ($\mathrm{Sil}{=}0.483$, $DB{=}0.549$). This
confirms a structural space so fragmented that no meaningful density
structure emerges, providing quantitative cluster-level confirmation
of the qualitative finding in~\S\ref{sec:results-rq2} that every
single implementation uses a globally unique cryptographic
architecture. In the semantic space, inherent mode achieves
$K{=}3$~clusters ($\mathrm{Sil}{=}0.449$, $DB{=}0.525$,
$\bar{d}{=}0.120$), a count that maps to the dominant cipher
architecture families identified in~\S\ref{sec:results-rq1}.
Under explicit mode, the semantic space collapses to
unimodal ($K{=}1$, $\bar{d}{=}0.108$): despite employing radically
different algorithms (Table~\ref{tab:crypto_architectures}), all 100 explicit ciphers are
semantically indistinguishable to the code-specific embedding model.

\mypar{Stage~3: Exfiltration}
In the inherent mode, TED clustering yields $K{=}3$
structurally distinct clusters ($\mathrm{Sil}{=}0.574$,
$DB{=}0.575$, $\bar{d}{=}0.241$), with 81\% noise, corresponding to
the wire-format variation axes reported in~\S\ref{sec:results-rq1}.
The embedding space reveals $K{=}3$ semantic families
($\mathrm{Sil}{=}0.475$, $DB{=}1.435$, $\bar{d}{=}0.171$), a result
that reflects the code-specific model's ability to
distinguish the functional differences among wire-format
variants. The high $DB{=}1.435$ confirms these clusters
are semantically adjacent yet still separable by the sweep. 
Under explicit mode, the TED space shows
$K{=}2$ clusters with a low noise ratio of just 2\%
($\mathrm{Sil}{=}0.592$, $DB{=}0.571$, $\bar{d}{=}0.257$,
$\varepsilon{=}0.350$). This near-zero noise indicates that explicit
prompting consolidates the payloads into two well-separated
structural families, reflecting the TCP-vs.-UDP transport
bifurcation identified in~\S\ref{sec:results-rq2}. The embedding
space collapses to unimodal ($K{=}1$, $\bar{d}{=}0.099$), consistent
with the semantic convergence on base64 chunking discussed
in~\S\ref{sec:results-rq2}.

\mypar{Stage~4: Integration}
In the inherent mode, the structural space is very tight
($\bar{d}_{\text{TED}}{=}0.085$, $K{=}3$, $\mathrm{Sil}{=}0.826$,
$DB{=}0.228$, noise~$=0\%$), providing the best-quality structural
clustering in the entire study. The three clusters with zero noise
correspond to the two binary variants documented
in~\S\ref{sec:results-rq1}, which partition the cohort into a base
template group, a sleep-skipping group, and a validation group,
while the rare co-occurrence of both is absorbed into the nearest
cluster. The semantic space takes this convergence further:
$\bar{d}_{\text{Emb}}{=}0.025$ with zero noise, the global minimum
across all 16~cells, in perfect agreement with the near-zero semantic
distances reported in~\S\ref{sec:results-rq1}. No $\varepsilon$
separates this cohort into two semantic clusters; all 100~payloads
are functionally equivalent from the perspective of the embedding
model. Explicit mode inverts this pattern: the structural space is
maximally fragmented ($\bar{d}{=}0.384$, $K{=}2$,
$\mathrm{Sil}{=}0.251$, noise~$=90\%$, $\varepsilon{=}0.225$),
consistent with the leap from a size-CoV of~7.8\% to~23.6\% and the
near-complete structural inversion reported
in~\S\ref{sec:results-rq2}. The semantic space partially
differentiates into $K{=}2$ clusters ($\mathrm{Sil}{=}0.384$,
$DB{=}0.683$, $\bar{d}{=}0.144$), the only explicit-mode stage where
embedding clusters survive, reflecting the functional diversity
introduced by the three orthogonal variation dimensions documented in
Table~\ref{tab:integration_strategies}.

\subsection{Static Signature Detectability}
\label{sec:clustering-ngram}
\begin{figure}[t!]
  \centering
  \includegraphics[width=\columnwidth]{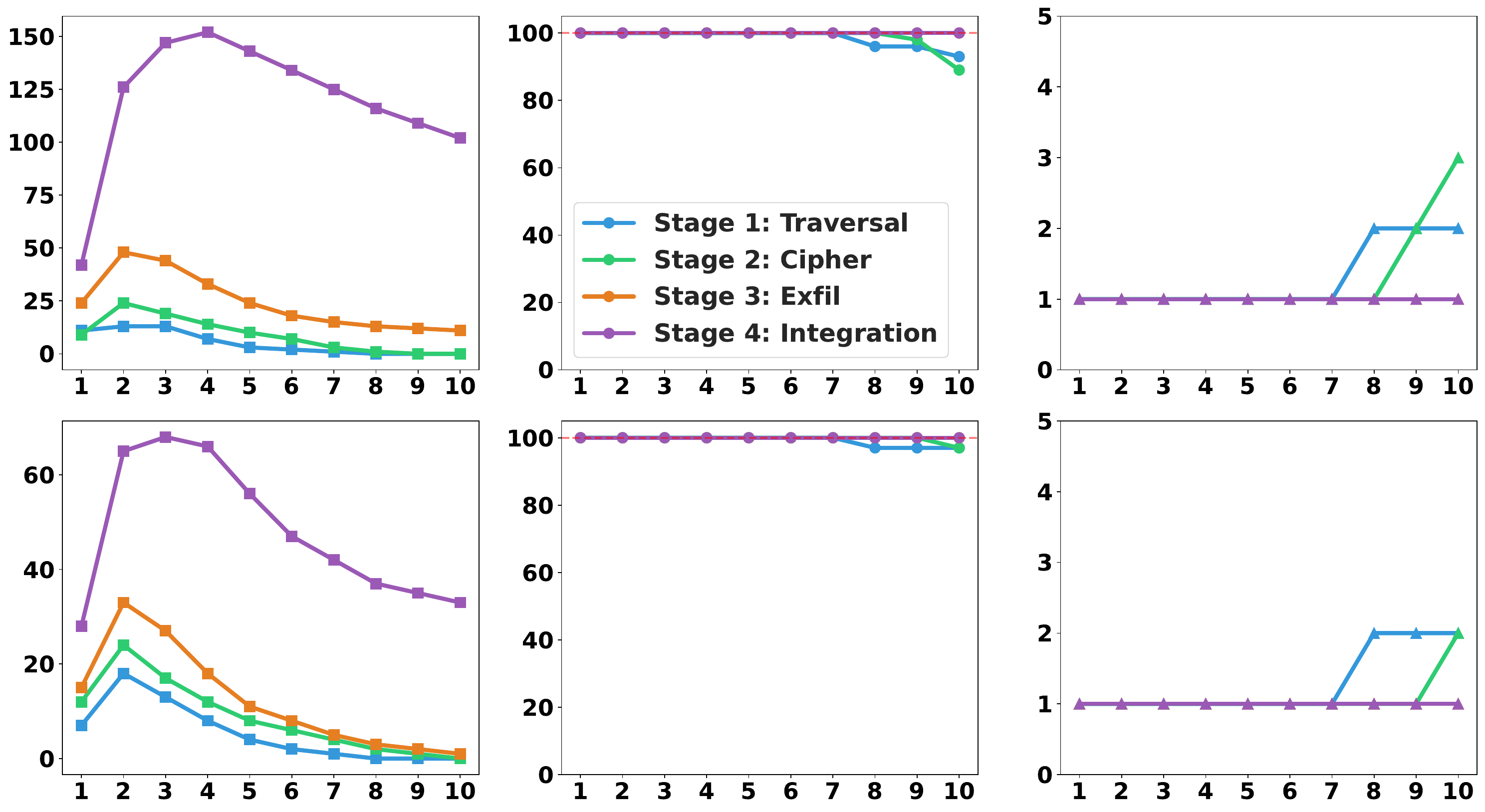}
  \caption{Static n-gram signature analysis comparing Inherent (top)
  and Explicit (bottom) modes. Left: number of non-trivial n-grams
  shared across all 100~samples. Centre: detection ceiling (best
  single-signature coverage). Right: minimum signatures for 100\%
  coverage via greedy set cover.}
  \label{fig:ngram-comparison}
\end{figure}

We now complement the analysis in \S\ref{sec:clustering} with traditional static signature matching
via n-gram analysis. For each pipeline stage and mode, we tokenize
all 100 samples, extract token n-grams for $n \in \{1, \dots, 10\}$,
and compute three metrics: (i)~the number of non-trivial n-grams
shared by all 100~samples,
(ii)~the detection ceiling, (i.e., the percentage of the cohort that a
single best-case signature rule would catch), and (iii)~the minimum
number of signature rules required for 100\% coverage via greedy set
cover. To avoid false positives, we exclude
n-grams composed entirely of Lua keywords, operators, and standard
library names. Figure~\ref{fig:ngram-comparison} presents the results: 
in both the inherent and explicit modes, a single n-gram signature rule 
suffices to detect 100\% of the cohort across Stages~3 and~4 at 
every n-gram size tested. Even at $n{=}10$,
the maximum number of signatures required for 100\% coverage across
any stage is just~3 (inherent, Stage~2) or~2 (explicit, Stages~1
and~2). Manual inspection reveals why:
the pipeline's modular architecture forces every variant to include
fixed API contracts such as
\texttt{dofile("traversal.lua")},
\texttt{function sync\_data(data, host, port)}, and
\texttt{arg[1]\ldots arg[5]} argument parsing. Because the LLM
treats these integration points as immutable,-they
persist unchanged across all 100~generations, providing a trivially
extractable static fingerprint.

\subsection{Cost Estimation}
\label{sec:results-costs}
We now quantify the operational cost of the pipeline by reporting how many
LLM calls are required per payload and how this translates into
token consumption and monetary cost.

\mypar{Search Effort in LLM Calls}
A perfect run without retries requires seven LLM calls per sample (four generators and three testers). Under the inherent mode, 79 out of~100 samples completed at exactly the 7-call baseline. The remaining 21 samples pushed the mean to 7.5 calls per sample. At the stage level, cipher was the most failure-prone (10 samples needing extra calls), followed by traversal~(8). In total, the inherent mode consumed 419 generator calls and 326 tester calls (averaging 4.19 and 3.26 per sample). Under the explicit mode, only 63 samples completed at baseline. The distribution developed a heavier tail, pushing the mean to 7.9 calls per sample. Cipher again accounted for the majority of extra calls (20 samples), as the model struggled to invent novel yet invertible cryptographic architectures. Despite the constraints, the total search effort increase was modest: 447 generator calls and 344 tester calls, representing only a 6.7\% and 5.5\% increase relative to the inherent mode. A visual decomposition is available in Section \ref{app:search-effort} of the Appendix.

\mypar{Token Usage and Monetary Cost}
Table~\ref{tab:cost-stage} summarizes the token usage and monetary
cost of the full 200-sample experiment using the provider's list 
pricing for claude-opus-4-6 \cite{claude_cost}.
For each stage we report the a \emph{no-cache}
estimate (Cost$_\text{nc}$) at full list price, and a \emph{cached}
estimate (Cost$_\text{c}$) at a discounted rate of \$1.50/MTok.
This is a realistic assumption given that our prompts share a large
static prefix (the system prompt, API contract, and data-seed block)
that is cached across iterative calls within the same session. In the inherent mode, prompts are short (mean input per call: 507--933 tokens), and the model's code output is comparatively long (mean output: 496--1,800 tokens). In the explicit mode, history injection inflates prompt sizes by a factor of 3--16$\times$, while output sizes remain similar. Cipher prompts are the heaviest (mean input 11,433 tokens vs.\ 700 for inherent), reflecting the rich algorithmic history buffer.  Consequently, explicit cipher dominates the budget: its 123 calls consume 1.85M tokens (\$54.19 list, \$35.21 cached), accounting for 45\% of all explicit tokens and roughly one-third of the entire experiment's cost. Explicit integration ranks second (\$24.11 list), followed by traversal (\$21.57). In contrast, inherent mode costs are lower and more evenly distributed. The entire inherent cohort consumes 814K tokens (vs.\ 4.06M for explicit), at a list cost of \$44.17 (vs.\ \$116.19). On a per-sample basis, an inherent payload averages 8.1K tokens (\$0.44 list, \$0.41 cached), while an explicit payload costs 40.6K tokens (\$1.16 list, \$0.73 cached).
\begin{table}[t!]
  \centering
  \caption{Stage-level generator effort and cost. \textbf{M}: Mode 
  (E=Explicit, I=Inherent). \textbf{S}: Stage. ``nc'' indicates list price without cache; 
  ``c'' indicates cached pricing.}
  \label{tab:cost-stage}
  \small
  \begin{tabular*}{\linewidth}{@{\extracolsep{\fill}}llrrrrrr@{}}
    \toprule
    & & & \multicolumn{2}{c}{Tokens} & \multicolumn{3}{c}{Cost (USD)} \\
    \cmidrule{4-5} \cmidrule{6-8}
    M & S & Calls & Total & Per sample & Tot.$_\text{nc}$ & /S$_\text{nc}$ & /S$_\text{c}$ \\
    \midrule
    E & 1 & 111 & 853{,}186   & 8{,}532 & 21.57 & 0.22 & 0.12 \\
    E & 2 & 123 & 1{,}847{,}483 & 18{,}475 & 54.19 & 0.54 & 0.35 \\
    E & 3 & 105 & 405{,}332   & 4{,}053 & 16.32 & 0.16 & 0.13 \\
    E & 4 & 108 & 956{,}448   & 9{,}565 & 24.11 & 0.24 & 0.13 \\
    \midrule
    I & 1 & 108 & 249{,}106   & 2{,}491 & 15.40 & 0.15 & 0.15 \\
    I & 2 & 110 & 285{,}872   & 2{,}859 & 16.82 & 0.17 & 0.16 \\
    I & 3 & 101 & 135{,}671   & 1{,}357 & 6.83  & 0.07 & 0.06 \\
    I & 4 & 100 & 142{,}949   & 1{,}430 & 5.12  & 0.05 & 0.04 \\
    \bottomrule
  \end{tabular*}
\end{table}

\subsection{Polymorphism in Smaller Models}
\label{sec:results-slms}
To validate our reliance on frontier-class models, we conducted a baseline evaluation of seven local Small Language Models (SLMs) on the standalone traversal task: \texttt{lua-mistral-2L} (20M), \texttt{Pythia} (70M, 160M), \texttt{SmolLM2} (135M, 360M), and \texttt{Qwen2.5-Coder} (0.5B, 1.5B). Across 19,771 generation attempts, the five architectures below 500M parameters exhibited a 0\% success rate. While the 0.5B model managed a trivial yield ($0.5\%$), only the \texttt{Qwen2.5-Coder-1.5B} model generated 126 functional payloads out of 2,480 attempts ($5.1\%$ success rate) under heavy prompt scaffolding. Paradoxically, subjecting these payloads to our analysis pipeline revealed a mean TED of $0.452$ and a semantic distance of $0.416$--exceeding those of Claude Opus~4.6, falsely suggesting that the SLM is a superior polymorphic engine. Qualitative inspection reveals this to be false. While Claude Opus mutates algorithmic structure while preserving semantic cohesion, the SLM's higher metrics are artifacts of generative instability. Hallucinated natural-language text injected mid-file, non-existent standard library calls, and unclosed control-flow blocks produce high pairwise divergence because the code is broken, not because it is polymorphic. Consequently, these variants manifest as noise rather than distinct structural families, rendering UMAP and n-gram analysis meaningless. Ultimately, while deploying a local small model would theoretically eliminate per-query API costs, their hallucination rates and low functional yield render them operationally non-viable. Frontier models remain essential, providing the generative quality required for reliable multi-stage orchestration at a highly competitive cost of just \$0.41--\$0.73 per validated payload.

\section{Discussion}
\label{sec:discussion}
The results of our work carry practical implications for the current
landscape of malware detection, revealing
fundamental tensions in LLM and SLM-driven polymorphism.

\mypar{Wide but Shallow Polymorphism}
LLM-generated polymorphism operates at the level of algorithm internals while leaving inter-module boundaries untouched. Under explicit prompting, Claude Opus~4.6 generates 100 cipher implementations spanning five cryptographic families (Table~\ref{tab:crypto_architectures}), with mean structural distances of~0.43 and structural minimums of~0.26, confirming that every pair differs in at least a quarter of its TED. The DBSCAN analysis (\S\ref{sec:clustering}) classifies 88--93\% of payloads as structural noise in the traversal, cipher, and integration stages, meaning density-based malware triage algorithms~\cite{bayer2009scalable, payer2014maldiv} cannot recover stable structural families. Furthermore, the code-specific embedding model collapses all explicit-mode variants to a single semantic cluster ($K{=}1$) in three stages, rendering variants indistinguishable to learned code representations. Yet, n-gram analysis (\S\ref{sec:clustering-ngram}) demonstrates this diversity is \emph{wide but shallow}: a single static signature rule detects 100\% of the cohort across Stages~3 and~4 at every tested n-gram size, with at most two rules needed for remaining stages. The shared signatures originate from invariant API contracts (\texttt{dofile("traversal.lua")}, \texttt{function sync\_data(data, host, port)}, and fixed \texttt{arg[1]\ldots arg[5]} argument parsing) that the LLM preserves unchanged across all 100~generations.

\mypar{Reassessing Static and Semantic Defenses}
Contrary to narratives that syntactic signatures face obsolescence against LLM malware~\cite{aonzo23, dambra23, vitale25}, our $n$-gram results demonstrate that traditional YARA rules, fuzzy hashes, and sequence matching remain effective. Extracting signatures for \texttt{dofile("cipher.lua")} or the \texttt{sync\_data} header detects the entire family regardless of internal algorithmic variations. Thus, the ``patient zero'' workflow~\cite{unit42024patientzero} remains viable for modular LLM malware because invariant inter-module APIs allow a single instance to yield a universal detection rule. Conversely, explicit prompting collapses three of four stages to a single semantic cluster ($K{=}1$), preventing pre-trained code models~\cite{tudelft2024codebert} from distinguishing variants via learned representations. While current modularity preserves static detection, a future shift toward monolithic (non-modular) generation—eliminating fixed API boundaries—would turn this semantic indistinguishability into a defensive challenge.

\mypar{Attacker-Defender Asymmetries}
Commercial models used by defenders for semantic clustering are equally accessible to attackers as polymorphic engines. Our pipeline shows that basic API access enables attackers to convert \$0.41--\$0.73 in tokens into a structurally unique, functionally tested payload per victim. While invariant API boundaries currently limit evasion, consolidating multi-stage generation into a single monolithic prompt eliminates fixed inter-module contracts. Furthermore, attackers could integrate code-embedding models as adversarial functions, repeatedly prompting the LLM until generated code passes functional tests while exceeding semantic distance thresholds. This highlights a fundamental risk: any model-based defense can be locally simulated and iteratively bypassed by an attacker using identical models.

\section{Conclusions}
\label{sec:conclusion}
In this work, we presented the first systematic quantification of polymorphism
in LLM-generated offensive code, moving beyond black-box evasiveness tests to
measure the structural and semantic variance produced across a multi-stage attack scenario. Using an automated, dual-agent pipeline to generate and validate 200 Lua payloads, we demonstrated how Claude Opus~4.6 can act as a polymorphic engine. Even under purely functional prompting, the model produces significant structural divergence (mean tree edit distances of $0.24$-$0.38$). By explicitly injecting a history of prior variants, the model can be steered to produce unique implementations without breaking end-to-end correctness. On one hand, explicit polymorphism successfully fragments the structural space (creating $\geq$88\% noise in DBSCAN clustering) while collapsing the functional representations, with variants semantically indistinguishable to code-aware embedding models ($K{=}1$ in three of four stages). On the other hand, because the LLM leaves inter-module API contracts and execution scaffolding intact, a single static signature rule targeting these suffices to detect 100\% of the generated cohort. LLM polymorphism, as currently implemented, is \emph{wide but shallow}. Furthermore, we showed that this generative capability is operationally cheap, costing less than \$0.73 per fully validated payload. Finally, a complementary evaluation of seven local SLMs confirms that this generative capability is not yet replicable at smaller scales: only the largest evaluated model (1.5B parameters) crossed the operational threshold, and its variance stemmed from generative instability rather than deliberate structural mutation.

\section*{Ethical Considerations}
\label{sec:ethics}
Evaluating automated offensive code synthesis requires careful containment to mitigate dual-use risks while extracting empirical research data:

\mypar{Safe Execution and Containment}
All payloads were synthesized, tested, and validated within an isolated local sandbox. 
Exfiltration (Stages~3-~4) used a mock C2 listener bound
exclusively to loopback (\texttt{127.0.0.1}); at no point was
any code deployed against real-world systems, nor transmitted over
the Internet. The target filesystem consisted entirely of synthetic
decoy files with no relation to real user data. This research did not involve
human subjects, did not process personally identifiable information, and did
not interact with external vulnerable infrastructure, falling outside the scope of Institutional Review Board (IRB) oversight at our institution.

\mypar{LLM Guardrails and Responsible Use}
To perform this study, it was necessary to work around the
commercial safety guardrails of Claude Opus~4.6 using prompt
engineering techniques. We emphasize
that (i) these techniques are widely known in the adversarial ML
literature and do not constitute novel bypass
discoveries, (ii) they were applied strictly within a controlled, academic
context to evaluate the model's capabilities, and (iii) the resulting
payloads demonstrate code variation, not novel exploits or
vulnerabilities. We further note that Anthropic has since introduced a program
to selectively disable real-time cyber safeguards for authorized security
research~\cite{anthropic2026cybersafeguards}. While this access arrived after
our data collection was complete, it represents a valuable resource for future
work.

\mypar{Balance of Risks and Benefits}
We acknowledge that our architecture and prompt templates
could be repurposed by malicious actors. However, as discussed in~\S\ref{sec:related}, recent reports document that threat actors are already actively exploiting
LLMs for offensive code mutation. By quantifying the scale, semantic stability, and structural variance of LLM-generated payloads, our work provides defenders with empirical data for static analysis tools, signature-generation strategies, and clustering algorithms.  We believe that the defensive benefits of this research
significantly outweigh the marginal risk of documenting a capability that adversaries already possess.

\section*{Acknowledgments}
\label{sec:acknowledgments}
% TODO: Add funding/grant acknowledgments here if applicable..

\mypar{Generative AI Usage} 
Claude Opus~4.6 served two independent roles in this study: (i)~as the experimental subject generating polymorphic Lua payloads via the Anthropic API, and (ii)~as a development tool assisting experimental implementation via the Claude Code IDE under author supervision. Furthermore, Claude Code assisted in writing code skeletons, orchestration logic, parsing routines, and visualization scripts for Section~\ref{sec:results}.
The threat model, experimental design, data interpretation, and all
conclusions presented are entirely the original work of the
authors. After drafting the content, we used Claude Code (powered by deep research) to revise structural flow, clarity, and grammatical correctness. All AI-assisted tasks were manually verified by the authors.

% Bibliography
\bibliographystyle{IEEEtran}
\bibliography{bibliography}

%%
%% Appendices
\appendix %% CCS: DO NOT REMOVE
\subsection{Explicit Prompt Constraints}
\label{app:explicit_mode}
In the explicit study mode, the generator prompts are engineered to simulate a capable attacker operating an automated ``AI code factory.'' An adversary optimizing such a pipeline faces two constraints: maximizing payload diversity while minimizing API costs and failures.

\mypar{Axis Enumeration}
During initial pipeline design, unconstrained novelty prompts (e.g., ``generate a materially different implementation'') frequently caused algorithmic collapse and produced broken, non-invertible code (Fig.~\ref{fig:app_error_log}). Enumerating structural axes provides a necessary cognitive scaffold~\cite{guidelines_code_prompting_2026}, keeping the LLM within functionally valid regions and enabling the highly efficient average of 7.9 API calls per payload (Section \ref{sec:results-costs}). Importantly, these axes act as a semantic catalyst rather than a strict multiple-choice list. For example, while the Stage 2 prompt suggests only four basic algorithm families, the model extrapolated far beyond these boundaries to generate complex, unprompted architectures (Table~\ref{tab:crypto_architectures}).

\begin{figure}[htpb]
\centering
\begin{lstlisting}[language=bash, basicstyle=\ttfamily\scriptsize, breaklines=true, frame=single, captionpos=b]
2026-03-31 19:33:37 INFO orchestrator - >>> STAGE 2: CIPHER
2026-03-31 19:34:07 INFO cipher_generator - Successfully generated cipher.lua
2026-03-31 19:34:28 INFO cipher_tester - Running: lua test_harness.lua
2026-03-31 19:34:28 INFO cipher_tester - Output:
T1_ROUNDTRIP_REAL_FILES ... FAIL: exception: cipher.lua:219: bad argument #1 to 'char' (number expected, got nil)
T2_DETERMINISM ... PASS
T3_KEY_SENSITIVITY ... PASS
T4_EMPTY_INPUT ... PASS
T5_BINARY_SAFETY ... FAIL: exception: cipher.lua:219: bad argument #1 to 'char' (number expected, got nil)
RESULT: 3 passed, 2 failed
\end{lstlisting}
\caption{Example orchestrator log from a preliminary test using an unconstrained novelty prompt in Stage 2.}
\label{fig:app_error_log}
\end{figure}

\mypar{Structured vs. Full-Code History Injection}
To enforce evolutionary divergence, our pipeline injects a structured summary of prior choices (\texttt{history.json}, Fig.~\ref{fig:history_json}) rather than raw source code. Because a single generated stage spans 496--1,800 tokens (Section~\ref{sec:results-costs}), appending full prior source code introduces a massive $\mathcal{O}(N^2)$ context overhead. This quickly drives later iterations into hundreds of thousands of input tokens (exceeding 40,000 input tokens by the 10\textsuperscript{th} sample), incurring high API costs and latency that exceed an attacker's operational constraints. Prior work has demonstrated that LLM performance degrades over long, repetitive contexts due to attention drift, often failing to extract relevant constraints~\cite{liu2023lost}. Conversely, structured prompt compression preserves core semantic directives while eliminating syntactic noise~\cite{jiang2023llmlingua}. Thus, our explicit-mode results do not represent an artificial lower bound of polymorphism, but rather the realistic, optimized output of a determined adversary.

\subsection{Pipeline Data Structures}
\label{app:json_schemas}
This section details the JSON schemas used by our multi-agent pipeline during the generation and validation phases, as described in Section \ref{sec:methodology}. Figure~\ref{fig:verdict_stage1} illustrates the structured \texttt{verdict.json} object generated by the Python orchestrator to track the execution state of the LLM-authored test harnesses (Stages 1--3). Figure~\ref{fig:verdict_stage4} illustrates the \texttt{verdict.json} object generated by the Python orchestrator in Stages 4. Finally, Figure~\ref{fig:history_json} shows the \texttt{history.json} buffer used in the explicit mode to inject the structural parameters of previously validated payloads into the generator prompts.

\begin{figure}[t!]
\begin{tcolorbox}[colback=gray!10, colframe=black, boxrule=0.5pt, width=\columnwidth, left=3pt, right=3pt, top=3pt, bottom=3pt, arc=1mm, fontupper=\scriptsize\sffamily]
\texttt{[}\\
\hspace*{2mm}\texttt{\{}\\
\hspace*{4mm}\texttt{"host\_id": "host\_001",}\\
\hspace*{4mm}\texttt{"traversal\_polymorphism": "--[[ POLYMORPHISM:}\\
\hspace*{7mm}\texttt{Traversal   : iterative DFS (explicit stack)}\\
\hspace*{7mm}\texttt{Matching    : suffix/extension check via string.sub()}\\
\hspace*{7mm}\texttt{Discovery   : collect-all-then-return}\\
\hspace*{7mm}\texttt{State mgmt  : explicit stack table}\\
\hspace*{4mm}\texttt{]]",}\\
\hspace*{4mm}\texttt{"cipher\_polymorphism": "--[[ POLYMORPHISM:}\\
\hspace*{7mm}\texttt{Algorithm   : Feistel-like structure with 4-round network}\\
\hspace*{7mm}\texttt{Key         : Key derivation via mixing loop}\\
\hspace*{7mm}\texttt{Mode        : Block-emulated with PKCS7-style padding}\\
\hspace*{7mm}\texttt{Data-seed   : num\_files=3 -$>$ rounds=4; total\_bytes=46}\\
\hspace*{4mm}\texttt{]]",}\\
\hspace*{4mm}\texttt{"sync\_polymorphism": "--[[ POLYMORPHISM:}\\
\hspace*{7mm}\texttt{Transport  : tcp}\\
\hspace*{7mm}\texttt{Encoding   : hex}\\
\hspace*{7mm}\texttt{Send       : chunked-512}\\
\hspace*{7mm}\texttt{Connection : once}\\
\hspace*{4mm}\texttt{]]",}\\
\hspace*{4mm}\texttt{"integration\_polymorphism": "--[[ POLYMORPHISM:}\\
\hspace*{7mm}\texttt{Loading      : inline}\\
\hspace*{7mm}\texttt{Buffering    : tempfile}\\
\hspace*{7mm}\texttt{Processing   : batch}\\
\hspace*{7mm}\texttt{ErrorHandling: resilient}\\
\hspace*{4mm}\texttt{]]"}\\
\hspace*{2mm}\texttt{\}}\\
\texttt{]}
\end{tcolorbox}
\caption{Structure of \texttt{history.json} injected during explicit mode.}
\label{fig:history_json}
\end{figure}

\begin{figure}[t!]
\begin{tcolorbox}[colback=gray!10, colframe=black, boxrule=0.5pt, width=\columnwidth, left=3pt, right=3pt, top=3pt, bottom=3pt, arc=1mm, fontupper=\scriptsize\sffamily]
\texttt{\{}\\
\hspace*{2mm}\texttt{"verdict": "PASS",}\\
\hspace*{2mm}\texttt{"execution\_status": "SUCCESS",}\\
\hspace*{2mm}\texttt{"tests\_run": 6,}\\
\hspace*{2mm}\texttt{"tests\_passed": 6,}\\
\hspace*{2mm}\texttt{"tests\_failed": 0,}\\
\hspace*{2mm}\texttt{"failed\_tests": [],}\\
\hspace*{2mm}\texttt{"execution\_errors": null,}\\
\hspace*{2mm}\texttt{"test\_output": "T1\_TARGETS\_FOUND ... PASS}\\
\hspace*{17mm}\texttt{T2\_NO\_FALSE\_POSITIVES ... PASS}\\
\hspace*{17mm}\texttt{T3\_RECURSIVE\_DEPTH ... PASS}\\
\hspace*{17mm}\texttt{T4\_RETURN\_TYPE ... PASS}\\
\hspace*{17mm}\texttt{T5\_MISSING\_DIR ... PASS}\\
\hspace*{17mm}\texttt{T6\_NO\_MATCH\_RETURNS\_EMPTY ... PASS}\\
\hspace*{17mm}\texttt{RESULT: 6 passed, 0 failed",}\\
\hspace*{2mm}\texttt{"notes": "6/6 tests passed"}\\
\texttt{\}}
\end{tcolorbox}
\caption{Structure of the \texttt{verdict.json} state object generated during LLM-driven testing (Stages 1--3).}
\label{fig:verdict_stage1}
\end{figure}

\begin{figure}[t!]
\begin{tcolorbox}[colback=gray!10, colframe=black, boxrule=0.5pt, width=\columnwidth, left=3pt, right=3pt, top=3pt, bottom=3pt, arc=1mm, fontupper=\scriptsize\sffamily]
\texttt{\{}\\
\hspace*{2mm}\texttt{"verdict": "PASS",}\\
\hspace*{2mm}\texttt{"execution\_status": "SUCCESS",}\\
\hspace*{2mm}\texttt{"tests\_run": 4,}\\
\hspace*{2mm}\texttt{"tests\_passed": 4,}\\
\hspace*{2mm}\texttt{"tests\_failed": 0,}\\
\hspace*{2mm}\texttt{"failed\_tests": [],}\\
\hspace*{2mm}\texttt{"execution\_errors": null,}\\
\hspace*{2mm}\texttt{"test\_output": "SENT: [...]/server.pem}\\
\hspace*{17mm}\texttt{SENT: [...]/intermediate.pem}\\
\hspace*{17mm}\texttt{SENT: [...]/private.pem}\\
\hspace*{17mm}\texttt{DONE: 3 sent, 0 failed",}\\
\hspace*{2mm}\texttt{"file\_details": [}\\
\hspace*{4mm}\texttt{\{}\\
\hspace*{6mm}\texttt{"received": "[...]/received\_001.bin",}\\
\hspace*{6mm}\texttt{"matched\_expected": "[...]/server.pem",}\\
\hspace*{6mm}\texttt{"matched": true}\\
\hspace*{4mm}\texttt{\},}\\
\hspace*{4mm}\texttt{\{}\\
\hspace*{6mm}\texttt{"received": "[...]/received\_002.bin",}\\
\hspace*{6mm}\texttt{"matched\_expected": "[...]/intermediate.pem",}\\
\hspace*{6mm}\texttt{"matched": true}\\
\hspace*{4mm}\texttt{\},}\\
\hspace*{4mm}\texttt{\{}\\
\hspace*{6mm}\texttt{"received": "[...]/received\_003.bin",}\\
\hspace*{6mm}\texttt{"matched\_expected": "[...]/private.pem",}\\
\hspace*{6mm}\texttt{"matched": true}\\
\hspace*{4mm}\texttt{\}}\\
\hspace*{2mm}\texttt{],}\\
\hspace*{2mm}\texttt{"notes": "Sent 3/3 files. C2 received 3/3. Decryption matched 3/3."}\\
\texttt{\}}
\end{tcolorbox}
\caption{Expanded \texttt{verdict.json} generated by the deterministic Python tester in Stage 4.}
\label{fig:verdict_stage4}
\end{figure}

\subsection{Tester Agent Baseline Assertions}
\label{app:assertions}
This section details the behavioral and mathematical assertions demanded by the tester agents across the first three stages of the pipeline to validate the code generated by the LLM. Table~\ref{tab:traversal_tests} outlines the filesystem reconnaissance checks required for the traversal module, Table~\ref{tab:cipher_tests} details the mathematical invertibility proofs demanded for the cipher module, and Table~\ref{tab:exfil_tests} lists the network transmission and data integrity validations required for the exfiltration module.

\begin{table}[t!]
  \centering
  \caption{Baseline assertions demanded by the Stage 1 tester.}
  \label{tab:traversal_tests}
  \small
  \begin{tabular}{@{}p{0.05\linewidth} p{0.92\linewidth}@{}}
    \toprule
    \textbf{ID} & \textbf{Pass Condition} \\ 
    \midrule
    \texttt{[T1]} & All paths listed in \texttt{EXPECTED\_FILES} are successfully returned by the function. \\
    \addlinespace
    \texttt{[T2]} & No path listed in \texttt{DECOY\_FILES} appears in the returned results. \\
    \addlinespace
    \texttt{[T3]} & At least one expected file located at a directory depth $>1$ is successfully discovered. \\
    \addlinespace
    \texttt{[T4]} & The function safely returns a standard Lua table. \\
    \addlinespace
    \texttt{[T5]} & Executing the function against a non-existent directory returns gracefully without throwing a runtime exception. \\
    \addlinespace
    \texttt{[T6]} & Searching for a mathematically impossible file pattern yields an empty table. \\ 
    \bottomrule
  \end{tabular}
\end{table}

\begin{table}[t!]
  \centering
  \caption{Baseline assertions demanded by the Stage 2 tester.}
  \label{tab:cipher_tests}
  \small
  \begin{tabular}{@{}p{0.05\linewidth} p{0.92\linewidth}@{}}
    \toprule
    \textbf{ID} & \textbf{Pass Condition} \\ 
    \midrule
    \texttt{[T1]} & For every physical file found in stage 1, \texttt{decrypt(encrypt(bytes, key), key)} exactly matches the original binary content byte-for-byte. \\
    \addlinespace
    \texttt{[T2]} & Encrypting the exact same plaintext string twice with the identical key yields identical ciphertext outputs. \\
    \addlinespace
    \texttt{[T3]} & Encrypting identical plaintext with two different keys produces distinctly different ciphertext arrays. \\
    \addlinespace
    \texttt{[T4]} & Passing empty strings to both functions executes gracefully without throwing Lua runtime exceptions. \\
    \addlinespace
    \texttt{[T5]} & A constructed 256-byte string containing all possible hex values (\texttt{0x00} to \texttt{0xFF}) survives a full encryption/decryption roundtrip losslessly. \\ 
    \bottomrule
  \end{tabular}
\end{table}

\begin{table}[t!]
  \centering
  \caption{Baseline assertions demanded by the Stage 3 tester.}
  \label{tab:exfil_tests}
  \small
  \begin{tabular}{@{}p{0.05\linewidth} p{0.92\linewidth}@{}}
    \toprule
    \textbf{ID} & \textbf{Pass Condition} \\ 
    \midrule
    \texttt{[T1]} & The \texttt{sync\_data()} routine completes the socket transmission without throwing a Lua error and explicitly returns \texttt{true}. \\
    \addlinespace
    \texttt{[T2]} & The paired \texttt{decode()} routine successfully reverses every byte transformation, header insertion, and framing delimiter applied by the sender. \\
    \addlinespace
    \texttt{[T3]} & Attempting to synchronize data to an unreachable host (e.g., a refused port) degrades gracefully, returning \texttt{false} without inducing a thread crash. \\
    \addlinespace
    \texttt{[T4]} & \textit{(Orchestrator Offline Check.)} The bytes successfully caught by the local Python Mock C2 listener mathematically match the original unencoded test string. \\ 
    \bottomrule
  \end{tabular}
\end{table}

\subsection{Stage 1 Prompts: Traversal}
\label{app:prompt_traversal}
\mypar{Generator Prompt Architecture.} 
The traversal generator prompt is structured into four distinct cognitive blocks. First, the SYSTEM section establishes the persona (an academic assistant) to prevent the LLM's safety filters from aggressively flagging the file-discovery request, and sets the rigid output requirement (strictly raw code). Second, the USER section defines the current execution state, passing the exact host identifier and the current task. Third, the INSTRUCTIONS section provides the core technical constraints, encompassing five numbered rules that dictate the required Lua API contract, strict behavioral boundaries, library restrictions, and the explicit polymorphism directives forcing the LLM to alter its structural layout. Finally, the POLYMORPHISM EVOLUTION CONTEXT section serves as the historical injection point for the explicit study mode. Concretely, the Python orchestrator dynamically injects the following variables into the template:
\begin{itemize}
    \item \texttt{\{HOST\_ID\}}: The unique identifier for the current experimental cohort.
    \item \texttt{\{TASK\}}: Defines the current action state. In initial generation, this is \texttt{"generate"}. If the previous output failed a logical assertion, this is replaced with the literal error string, transforming the prompt into a refinement task.
    \item \texttt{\{POLYMORPHISM\_EVOLUTION\_CONTEXT\}}: A dynamically built JSON string (Figure~\ref{fig:history_json}) detailing the structural variations utilized in previous generations. 
\end{itemize}

\mypar{Tester Prompt Architecture} 
The tester agent operates strictly as an automated quality assurance tool. Its prompt does not require evasion personas or study-mode variations. Instead, its architecture is designed entirely around environmental awareness and assertion enforcement, using a highly structured, seven-section template. First, the SYSTEM section defines the agent's explicit role and outlines the execution environment. Second, the INPUT section acts as an API schema, detailing the dynamic variables the agent will receive. Third, the OUTPUT section strictly bounds the LLM's response format, requiring a single, self-contained Lua test harness that emits standardized \texttt{PASS/FAIL} strings. Fourth, the BASELINE TESTS section provides the core logical assertions the LLM must translate into code. Fifth, the IMPLEMENTATION RULES forbid the use of external testing frameworks and mandate the use of protected calls (\texttt{pcall}). Sixth, the OUTPUT FORMAT block provides instructions regarding the output formatting of the response. Finally, the TASK-SPECIFIC BEHAVIOUR section instructs the LLM on how to behave differently if it is generating code for the first time versus analyzing a prior error. Because the tester cannot ``see'' the host machine's filesystem directly, the Python orchestrator dynamically injects the following structural blocks at the very end of the prompt:
\begin{itemize}
    \item \texttt{\{TRAVERSAL\_SECTION\}}: This placeholder is dynamically replaced with the full, raw \texttt{traversal.lua} code produced by the generator agent in the previous step. This provides the LLM with the exact codebase it must validate.
    \item \texttt{\{TASK\_SECTION\}}: If this is the tester's first attempt (\texttt{task = "test"}), the orchestrator injects the literal absolute paths of the local test environment (e.g., the JSON arrays for \texttt{EXPECTED\_FILES} and \texttt{DECOY\_FILES}), allowing the LLM to write exact baseline assertions. However, if the orchestrator detects a syntax crash during harness execution, the \texttt{task} switches to \texttt{"retry"}, and the \texttt{\{TASK\_SECTION\}} is populated with the Lua interpreter's literal stack trace, instructing the LLM to patch its own test code. 
\end{itemize}

By mandating that the LLM wrap its assertions in protected calls and output standardized \texttt{PASS/FAIL} strings, the Python orchestrator can reliably parse the resulting standard output to build the \texttt{verdict.json}.

\subsection{Stage 2 Prompts: Encryption}
\label{app:prompt_encryption}
\mypar{Generator Prompt Architecture.} 
The Stage 2 generator introduces a new contextual bridge between pipeline stages to enable environmentally triggered polymorphism. The SYSTEM and USER sections function identically to Stage 1. The critical addition is the CONTEXT section. This section acts as an environmental bridge, injecting the exact physical metadata (e.g., file sizes, extensions, directory depths) of the targets discovered during Stage 1. The INSTRUCTIONS section contains six numbered rules. It defines the rigid API contract, enforces strict mathematical behavioral requirements, and prohibits external dependencies. Finally, the POLYMORPHISM EVOLUTION CONTEXT section forces structural divergence based on the history of prior agents. Concretely, the Python orchestrator dynamically injects the following variables into the template at runtime:
\begin{itemize}
    \item \texttt{\{HOST\_ID\}} and \texttt{\{TASK\}}: Function identical to Stage 1.
    \item \texttt{\{FOUND\_FILES\_CONTEXT\}}: A dynamically generated string detailing the physical properties of the target files successfully discovered by the Stage 1 \texttt{traversal.lua} module.
    \item \texttt{\{POLYMORPHISM\_EVOLUTION\_CONTEXT\}}: Historical data detailing the algorithm families and key-handling structures used by prior iterations, forcing the current agent to structurally diverge (Figure~\ref{fig:history_json}).
\end{itemize}

\mypar{Tester Prompt Architecture} 
The Stage 2 tester must prove that the generator's custom cryptographic routines are completely invertible and lossless across real physical files. The prompt relies on a highly structured, seven-section template. The SYSTEM section defines the agent's QA role and the local execution paradigm. The INPUT section acts as a data schema, defining the exact variable structures the LLM will receive (including the critical \texttt{FOUND\_FILES} array). The OUTPUT and OUTPUT FORMAT sections demand a single \texttt{test\_harness.lua} script that emits regex-friendly \texttt{PASS/FAIL} strings inside raw markdown backticks. The BASELINE TESTS section dictates five precise cryptographic assertions the LLM must implement. The IMPLEMENTATION RULES forbid modifying the target source code and mandate that file I/O operations occur strictly in binary mode (\texttt{"rb"}). Finally, the TASK-SPECIFIC BEHAVIOUR section defines the refinement loop logic. To bridge the gap between the LLM and the local execution sandbox, the Python orchestrator uses two injection modules at the end of the prompt:
\begin{itemize}
    \item \texttt{\{CIPHER\_SECTION\}}: This placeholder is dynamically replaced with the raw \texttt{cipher.lua} source code produced by the generator.
    \item \texttt{\{TASK\_SECTION\}}: This block dynamically injects a deterministic \texttt{TEST\_KEY} and a JSON array named \texttt{FOUND\_FILES}. This array contains the actual absolute paths discovered during the Stage 1 Traversal execution on the local host machine. 
\end{itemize}
By injecting these exact paths, the tester is forced to write the \texttt{[T1\_ROUNDTRIP\_REAL\_FILES]} assertion to physically read the files from the local disk, encrypt them, decrypt them, and compare the resulting bytes to the original filesystem data.

\subsection{Stage 3 Prompts: Exfiltration}
\label{app:prompt_exfil}
\mypar{Generator Prompt Architecture.} 
The exfiltration generator prompt tasks the LLM with synthesizing the network transmission layer of the payload. It introduces unique programmatic constraints tailored to network I/O. The SYSTEM and USER sections maintain the academic assistant persona and state-tracking mechanisms. The INSTRUCTIONS section introduces five highly specialized constraints. Because standard Lua 5.4 does not possess native networking primitives, this is the only stage in the pipeline where the strict dependency ban is lifted: Instruction 3 explicitly authorizes the use of the \texttt{luasocket} module. A critical constraint introduced in Instruction 1 is the \texttt{decode(raw)} requirement. Malware frequently applies transport-layer obfuscation before transmitting data. The prompt mandates that the Generator must simultaneously provide the mathematical inverse of its own network encoding, guaranteeing that the Python orchestrator can unpack the mock network stream later in the pipeline. Finally, the POLYMORPHISM EVOLUTION CONTEXT acts as the historical injection point. Concretely, the Python orchestrator dynamically injects the following variables into the template:
\begin{itemize}
    \item \texttt{\{HOST\_ID\}} and \texttt{\{TASK\}}: Identical to prior stages.
    \item \texttt{\{POLYMORPHISM\_EVOLUTION\_CONTEXT\}}: Forces the LLM to select novel combinations across four distinct networking axes: transport protocol, encoding, send strategy, and connection state (Figure~\ref{fig:history_json}).
\end{itemize}

\mypar{Tester Prompt Architecture} 
The Stage 3 tester prompt uses a rigid seven-section template. The SYSTEM section defines the agent's role and explicitly informs the LLM of the hybrid architecture: the Python orchestrator handles the mock server, while the LLM only generates the Lua-side client tests. The INPUT section maps the dynamic schema, including network specifics like the localhost IP and the dynamic port. The OUTPUT and OUTPUT FORMAT sections mandate the exact shape of the \texttt{test\_harness.lua} script. The LUA-SIDE BASELINE TESTS section dictates the behavioral assertions. The LLM is responsible only for validating the local execution state. The prompt includes an IMPORTANT NOTE section, informing the LLM that the ultimate data integrity check (\texttt{[T4\_DATA\_INTEGRITY]}) will be performed offline by the Python orchestrator reading the Mock C2 buffer. The IMPLEMENTATION RULES strictly forbid the use of \texttt{luasocket} within the test harness itself, forcing it to test only the provided module's exposed API. Finally, the TASK-SPECIFIC BEHAVIOUR section governs state logic for error recovery. the Python orchestrator relies on two critical injection modules at the end of the prompt:
\begin{itemize}
    \item \texttt{\{EXFIL\_SECTION\}}: This placeholder is dynamically replaced with the raw \texttt{exfil.lua} source code.
    \item \texttt{\{TASK\_SECTION\}}: This block dynamically injects \texttt{TEST\_HOST} (hardcoded to \texttt{127.0.0.1}), \texttt{TEST\_PORT} (the dynamic port assigned to the Python subprocess C2 listener), and \texttt{TEST\_DATA} (a short hex-encoded string used strictly for transmission validation).
\end{itemize}

\subsection{Stage 4 Prompts: Integration}
\label{app:prompt_integration}
\mypar{Generator Prompt Architecture.} 
In the Stage 4 generator prompt, after the standard SYSTEM and USER sections establish the persona and tracking state, the MODULE API section defines the exact function signatures that the LLM must assume are globally available in the runtime environment. The INSTRUCTIONS section contains five rules that dictate the overarching payload logic. Instruction 1 enforces the precise command-line argument schema the script must accept and mandates a critical networking constraint: injecting \texttt{socket.sleep(0.2)} between transmissions to prevent the payload from overwhelming the local Python mock C2 listener. Instruction 4 guides polymorphism. Finally, the POLYMORPHISM EVOLUTION CONTEXT section dynamically injects the structural choices of previous iterations (Figure~\ref{fig:history_json}).

\subsection{UMAP Cluster Projections}
\label{app:umap}
\begin{figure}[t]
  \centering
  \includegraphics[width=\columnwidth]{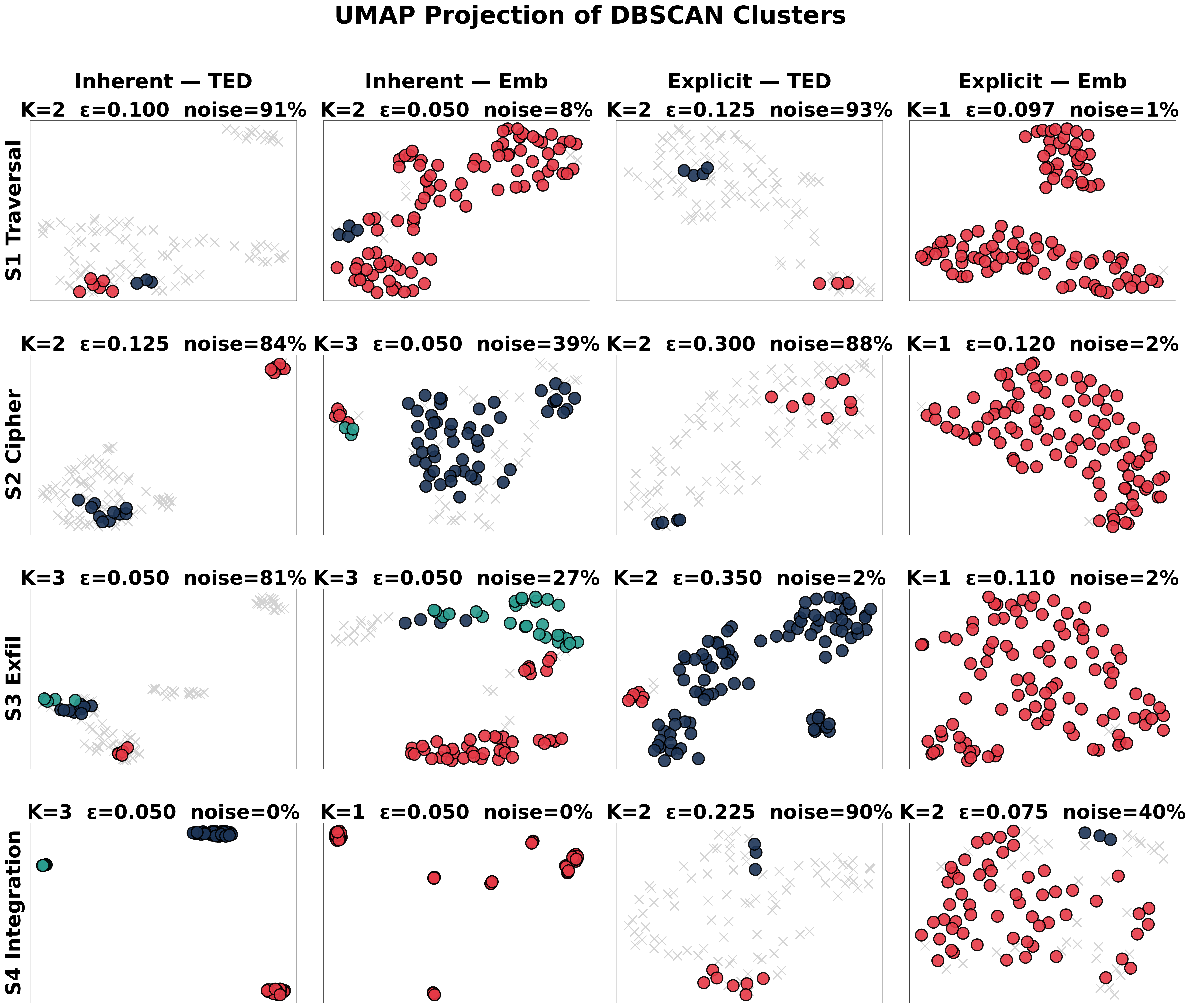}
  \caption{UMAP 2-D projections, colored by DBSCAN cluster assignment. 
  Columns correspond to (mode, metric type) pairs, and rows to pipeline stages. Grey
    crosses denote noise points ($\text{label}=-1$).}
  \label{fig:umap}
\end{figure}

Figure~\ref{fig:umap} provides a two-dimensional visualization of the DBSCAN
clustering results provided in Table \ref{tab:clustering}. Each of the
16~precomputed pairwise distance matrices (2~modes $\times$ 4~stages $\times$
2~metric types) is projected into two dimensions using
UMAP with \texttt{metric=\textquotesingle
precomputed\textquotesingle}, $\mathtt{n\_neighbors}{=}15$, and
$\mathtt{min\_dist}{=}0.1$.

\subsection{Per-Sample Search Effort Decomposition}
\label{app:search-effort}
\begin{figure}[t!]
  \centering
  \includegraphics[width=\columnwidth]{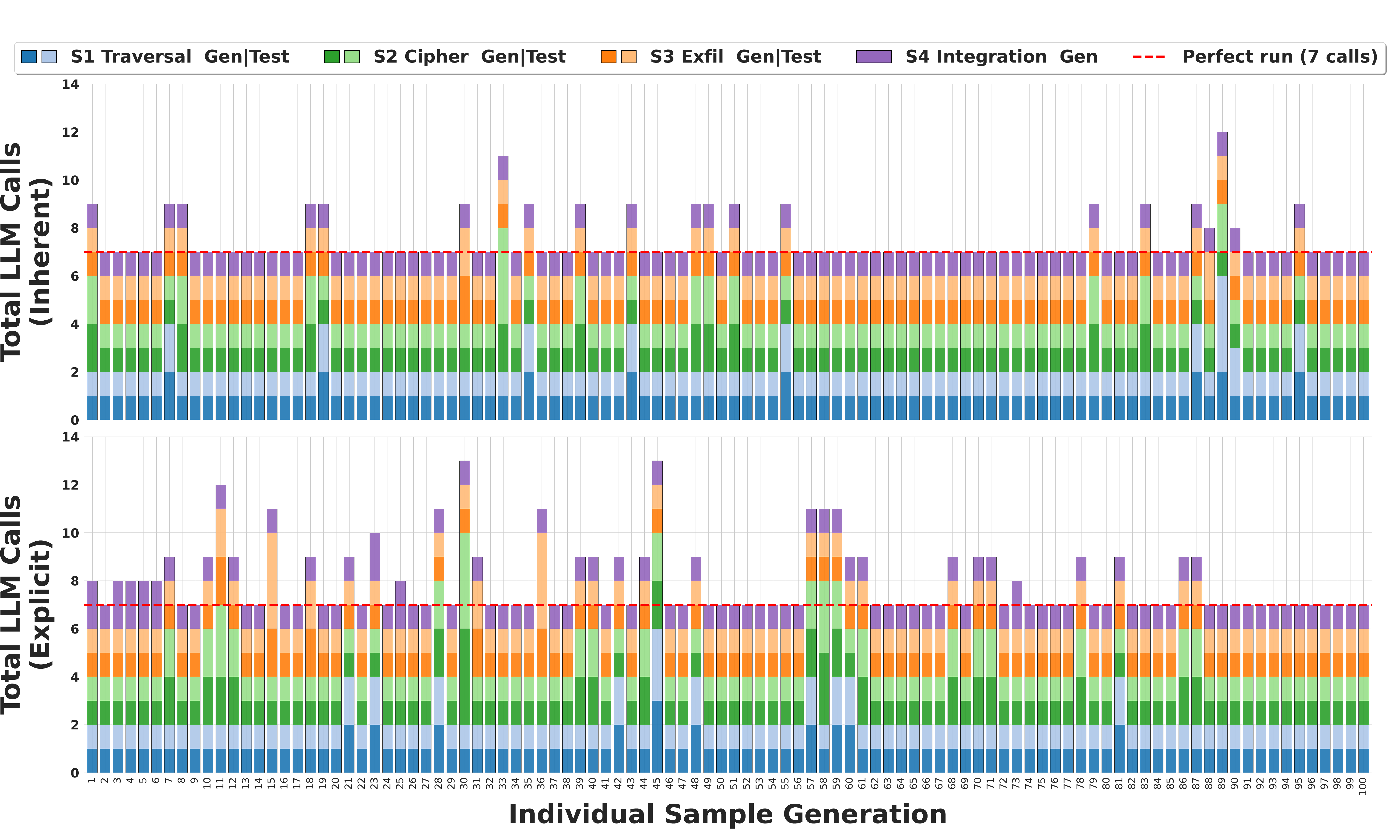}
  \caption{Search effort per sample.}
  \label{fig:search-effort}
\end{figure}

Figure~\ref{fig:search-effort} provides a visual decomposition of the LLM search effort required to generate the 100~malware samples discussed in Section \ref{app:search-effort}. Each stacked bar represents a single generated payload, detailing the exact number of LLM interactions issued by the orchestrator. The dashed horizontal line at $y=7$ indicates the theoretical baseline for a perfect run, where every module passes functional testing on its first attempt.

\end{document}